\documentclass[superscriptaddress,twocolumn,showpacs,prb,floatfix]{revtex4}
\usepackage{epsfig}

\begin{document}

\title{Critical properties of the three- and four-dimensional gauge glass}
\author{Helmut G.~Katzgraber}
\affiliation{Theoretische Physik, ETH H\"onggerberg,
CH-8093 Z\"urich, Switzerland}
\author{I.~A.~Campbell}
\affiliation{Laboratoire des Verres, Universit\'e Montpellier II, 34095
Montpellier, France}

\date{\today}

\begin{abstract}
The gauge glass in dimensions three and four is studied using a variety of
numerical methods in order to obtain accurate and reliable values for the
critical parameters. Detailed comparisons are made of the sensitivity of
the different techniques to corrections to finite-size scaling, which are
generally the major source of systematic error in such measurements.  For
completeness we also present results in two dimensions. The variation of
the critical exponents with space dimension is compared to results in
Ising spin glasses.
\end{abstract}

\pacs{75.50.Lk, 75.40.Mg, 05.50.+q}
\maketitle

\section{Introduction}
\label{introduction}

Spin glasses have long been recognized to be the canonical examples of
complex systems and so are of fundamental interest for statistical physics and
in particular the statistical physics of phase transitions. The critical
behavior at standard continuous transitions in systems {\it without}
disorder is now understood very precisely and in impressive detail thanks
to the renormalization group (RG) approach. An important step forward
towards the comprehension of the vast family of glassy transitions would
be to arrive at a similar degree of understanding concerning the critical
behavior at the spin-glass transition. Unfortunately this task has proven
to be extremely difficult. First of all, below the upper critical
dimension $(d = 6)$, the renormalization group theory that works so well
for standard transitions has been found to present formidable technical
difficulties\cite{dedominicis:98} and is of little guidance in lower
dimensions. Second, on the purely practical side, numerical work is
laborious and has tended to lead to imprecise values of ordering
temperatures and critical exponents. Randomly disordered samples are
microscopically inequivalent to each other and to obtain data that are
truly representative of the global average, measurements must be made over
large numbers of disorder realizations.  Time scales are long and
obtaining thermal equilibrium in large system sizes is hard. Finite-size
scaling methods are invariably used to estimate exponents but are carried
out on a restricted range of sizes as the need for extensive computer time
escalates when the system sizes increase. Because measurements are thus
intrinsically restricted to small or moderate system sizes, it is
essential to take into account possible artifacts due to corrections to
finite-size scaling. Little is known about the magnitude of these
corrections {\it a priori} and nevertheless the data must be analyzed in
such a way that existing corrections are identified and allowed for.

Among spin glasses the Ising (ISG) systems have been by far the most
studied numerically for obvious practical reasons. Here we report results
using a variety of numerical techniques on the gauge glass (GG) in space
dimensions $d = 3$ and $4$. This system is of interest in its own right
because vector spin systems are generally much closer to experimental
realizations of spin-glass ordering than are ISGs. Most laboratory spin
glasses are made up of Heisenberg spins with $O(3)$-symmetry and it has
been suggested that chiral ordering plays an important role in spin
freezing.\cite{kawamura:92,kawamura:98} Vortex glasses in type II
superconductors can be modeled by $XY$ vector spins [$O(2)$ symmetry], and
GGs have already been studied extensively in this
context.\cite{reger:91,reger:93,olson:00,katzgraber:02a} GGs are vector
spin glasses which, due to symmetry arguments, do not show chiral
ordering.\cite{olson:00}

This work has two main aims. The first is to take this particular family
of spin glasses as a case study in order to demonstrate that there exists
a whole toolbox of numerical techniques available to identify spin-glass
ordering transitions and to attempt to estimate as reliably and precisely
as possible the related critical exponents and the associated corrections
to scaling. Each of these techniques can have its advantages and
disadvantages, and we evaluate the reliability of the different methods.
Second, having values for the GG exponents in hand which are as precise as
possible, it is of interest to follow the evolution of their values as a
function of space dimension in this particular family and to compare with
values obtained in other families of spin glasses.

We have checked for and analyzed corrections to finite-size scaling in the
observables measured; the relative influence of these corrections varies
considerably from one observable to another. This can explain
inconsistencies between the estimates of critical temperatures and
critical exponents in various spin glasses which have been reported. We
conclude that the ``current'' defined later (related to the domain-wall
stiffness at low temperatures) is little affected by corrections to
finite-size scaling and thus provides reliable estimates of the ordering
temperature $T_{\rm c}$. Measurements using the spin-glass susceptibility
$\chi(L,T)$, alone or together with nonequilibrium measurements, are also
accurate and reliable. On the other hand in the systems with
finite-temperature transitions we have studied the intersection of the
correlation length divided by system size, $\xi_L(T)/L$, and find that the
method is very sensitive to corrections to finite-size scaling even for
sizes where these corrections have become negligible for other observables.
This method on its own would give imprecise or misleading estimates for
the ordering temperature of the GG. Nevertheless it can relatively well 
pinpoint if $T_{\rm c} > 0$, or not, in general.

In Sec.~\ref{model} we introduce the model studied. In Sec.~\ref{equil} we
discuss the equilibration test used in the (parallel tempering) Monte
Carlo method and describe the equilibrium observables measured. 
Off-equilibrium Monte Carlo methods and observables are discussed in 
Sec.~\ref{offequil}, followed by results in $d = 4$ in Sec.~\ref{4dgg} and
$d = 3$ in Sec.~\ref{3dgg}. Some results for $d = 2$ are presented in
Sec.~\ref{2dgg}. We conclude with a summary and a comparison of the techniques 
used to determine the critical parameters in Sec.~\ref{summary_comp}
and with a discussion about the dimensional dependence of the critical
exponents in Sec.~\ref{gg_exponents}. Concluding remarks are presented in
Sec.~\ref{summary}.

\section{Model}
\label{model}

The gauge glass is a canonical vector spin glass (see, for instance,
Ref.~\onlinecite{olson:00}) where $XY$ spins on a [hyper]cubic lattice of
size $L$ interact through the Hamiltonian
\begin{equation}
{\cal H} = -J \sum_{\langle i, j\rangle} \cos(\phi_i - \phi_j - A_{ij}) \; ,
\label{hamiltonian_gg}
\end{equation}
the sum ranging over nearest neighbors. The angles $\phi_i$ represent the
orientations of the $XY$ spins and the $A_{ij}$ are quenched random
variables uniformly distributed between $[0,2\pi]$ with the constraint
that $A_{ij} = -A_{ji}$.  $J$ is conventionally set equal to $1$. Periodic
boundary conditions are applied.

\section{Equilibrium Observables}
\label{equil}

Equilibrium measurements are carried out with samples fully thermalized
using the exchange Monte Carlo (parallel tempering)  
technique.\cite{hukushima:96,marinari:98b} We ensure equilibration by
checking that different observables do not change with the amount of Monte
Carlo steps and measure by doubling the number of Monte Carlo steps. Once
the last three measurements agree within error bars we are satisfied with
the equilibration.

For $XY$ spin systems there is a choice to be made in the allowed
single-spin acceptance angle for individual updating steps. To optimize
the updating procedure, the limiting angle is often chosen to be less than
$2\pi$ for an $XY$ spin\cite{katzgraber:01a} and linearly dependent on
temperature. The numerical prefactor for the temperature-dependent window
is chosen so that the acceptance ratios for the local Monte Carlo updates
is $\sim 0.4$. As far as the final equilibrium parameters are concerned,
this choice plays no role. However, for the nonequilibrium simulations to
be introduced later on it is important to use the full $2\pi$ acceptance
angle.

In Table~\ref{simparams_4d}, we show $N_{\mathrm{samp}}$ (number of
samples), $N_{\mathrm{sweep}}$ (total number of sweeps performed by each
set of spins), and $N_T$ (number of temperature values), used in the
simulations in four dimensions. Table~\ref{simparams_3d} has the
corresponding values for the simulations in three dimensions. The
parameters for the simulations in two dimensions are presented in
Table~\ref{simparams_2d} (see also Ref.~\onlinecite{katzgraber:03a}).

\begin{table}
\caption{
Parameters of the equilibrium simulations in four dimensions.
$N_{\mathrm{samp}}$ is the number of samples, $N_{\mathrm{sweep}}$ is the
total number of Monte Carlo sweeps for each of the $2 N_T$ copies 
(two replicas per temperature) for a
single sample, and $N_T$ is the number of temperatures used in the
parallel tempering method. $N_{\mathrm{samp}}(\xi_L)$ is the total number
of of disorder realizations used in the calculation of the two-point
correlation length $\xi_L$ (defined below). The lowest temperature used is
0.70, the highest 1.345.
\label{simparams_4d}
}
\begin{tabular*}{\columnwidth}{@{\extracolsep{\fill}} c c c c l }
\hline
\hline
$L$  &  $N_{\mathrm{samp}}$  & $N_{\mathrm{samp}}(\xi_L)$ &
$N_{\mathrm{sweep}}$ & $N_T$  \\
\hline
3 & $5000$ & $1660$ & $2.0 \times 10^4$ & 17 \\
4 & $5000$ & $1250$ & $8.0 \times 10^4$ & 17 \\
5 & $5000$ & $1000$ & $4.0 \times 10^5$ & 17 \\
\hline
\hline
\end{tabular*}
\end{table}

\begin{table}
\caption{
Parameters of the equilibrium simulations in three dimensions. The lowest
temperature simulated is 0.05, the highest 0.947. The different quantities
are explained in the caption of Table \ref{simparams_4d}.
\label{simparams_3d}
}
\begin{tabular*}{\columnwidth}{@{\extracolsep{\fill}} c c c c l }
\hline
\hline
$L$  &  $N_{\mathrm{samp}}$  & $N_{\mathrm{samp}}(\xi_L)$ & $N_{\mathrm{sweep}}$ & $N_T$  \\
\hline
3 & $10000$ & $2660$ & $6.0 \times 10^3$ & 53 \\
4 & $10000$ & $2000$ & $2.0 \times 10^4$ & 53 \\
5 & $10000$ & $1600$ & $6.0 \times 10^4$ & 53 \\
6 & $5000$  & $1330$ & $2.0 \times 10^5$ & 53 \\
8 & $2000$  & $1000$ & $1.2 \times 10^6$ & 53 \\
\hline
\hline
\end{tabular*}
\end{table}

\begin{table}
\caption{
Parameters of the equilibrium simulations in two dimensions. The lowest
temperature used is 0.13, the highest 1.058. For $L = 24$ the lowest
temperature studied is $0.20$. The different quantities are explained in
the caption of Table \ref{simparams_4d}.
\label{simparams_2d}
}
\begin{tabular*}{\columnwidth}{@{\extracolsep{\fill}} c c c c l }
\hline
\hline
$L$  &  $N_{\mathrm{samp}}$  & $N_{\mathrm{samp}}(\xi_L)$ &
$N_{\mathrm{sweep}}$ & $N_T$  \\
\hline
4  & 10400 & $4000$ & $8.0 \times 10^4$ & 30 \\
6  & 10150 & $2660$ & $8.0 \times 10^4$ & 30 \\
8  &  8495 & $2000$ & $2.0 \times 10^5$ & 30 \\
12 &  6890 & $1330$ & $8.0 \times 10^5$ & 30 \\
16 &  2500 & $1000$ & $2.0 \times 10^6$ & 30 \\
24 &  2166 &        & $2.0 \times 10^6$ & 24 \\
\hline
\hline
\end{tabular*}
\end{table}

A primary observable for a spin glass is the equilibrium spin-glass
susceptibility at finite system size $L$, $\chi(L,T)$.  The susceptibility
is defined as\cite{katzgraber:02a}
\begin{equation}
\chi = N[\langle q^2\rangle]_{\rm av} \; ,
\label{chisg}
\end{equation}
where $q$ is the spin-glass order parameter:
\begin{equation}
q = \frac{1}{N} \sum_{i = 1}^N \exp[i(\phi_i^\alpha - \phi_i^\beta)] \; .
\label{spinoverlap}
\end{equation}
Here, $\alpha$ and $\beta$ are two replicas of the system with the same
disorder and $[\cdots]_{\rm av}$ represents a disorder average, whereas
$\langle \cdots \rangle$ represents a thermal average; $N$ is the number
of spins in the system. The standard finite-size scaling form for the
equilibrium spin-glass susceptibility is
\begin{equation}
\chi(L) = L^{2-\eta} \tilde{C}[L^{1/\nu}(T-T_{\rm c})] \; .
\label{chi_eqn}
\end{equation}
The function $\tilde{C}$ tends to a constant at the critical temperature so
that at $T_{\rm c}$
\begin{equation}
\chi(L) \propto L^{2-\eta} \; ,
\label{chi_eqn_2}
\end{equation}
where $\eta$ and $\nu$ are the usual critical exponents.  Following the RG
approach, the spin-glass susceptibility, like the other observables,
will be modified at small $L$ by a correction to scaling factor $\sim [1+
AL^{-\omega}+ \cdots]$, where $\omega$ is the correction exponent arising
from the leading irrelevant operator in the RG and $A$ is a constant.  
However, in general there are also ``lattice artifact'' correction
terms\cite{salas:00} giving at $T_{\rm c}$
\begin{equation}
\chi(L) \propto L^{2-\eta}+B +\cdots \; ,
\label{chi_eqn_2.2}
\end{equation}
where $B$ has no relation to the RG correction. This leads to an effective
leading correction factor $[1+B^\prime L^{-(2-\eta)}]$. If $\omega >
(2-\eta)$ corrections will be dominated by the leading ``analytic'' term
with an effective correction exponent
$(2-\eta)$.\cite{salas:98,salas:00,ballesteros:99}

Spin-glass susceptibility measurements can be used directly to estimate
$T_{\rm c}$. As $\chi(L)$ in absence of corrections to scaling increases
as $L^{2-\eta}$ at the ordering temperature, a log-log plot\cite{log} 
of $\chi(L)$ against $L$ is linear at $T_{\rm c}$. At higher temperatures the 
plot will curve downward and at lower temperatures it will curve upward. The
crossover from negative to positive curvature at large $L$ should give a
precise measure of $T_{\rm c}$.  Alternatively one can plot
$\ln[\chi(nL)/\chi(L)]$ as functions of $T$ for different $L$, and, for
instance, $n=2$. The curves will intersect at $T_{\rm c}$.  Direct scaling of
$\chi(L,T)$ according to Eq.~(\ref{chi_eqn}) on the contrary provides a
very poor indication for $T_{\rm c}$.\cite{katzgraber:02a} In the presence of
corrections there will be an additional curvature at small $L$ which
should be essentially temperature independent near $T_{\rm c}$. If data on
a reasonably wide range of $L$ are available it is possible to identify
the leading correction term and to estimate $T_{\rm c}$ using data at
$L$ values large enough for the correction term to be negligible.

The Binder ratio\cite{binder:81} for the GG is defined by\cite{kawamura:00}
\begin{equation}
g(T)= 2-\frac{[\langle q^4\rangle]_{\rm av}}{[\langle q^2 
\rangle]_{\rm av}^2} \; .
\label{binder}
\end{equation}
Finite-size scaling predicts
\begin{equation}
g(T,L) = {\tilde G}[L^{1/\nu}(T - T_{\rm c})] \; .
\end{equation}
This dimensionless observable should be independent of $L$ at $T_{\rm c}$,
except for corrections to scaling. The Binder ratio is a {\em bona fide}
``work horse'' widely used to obtain an estimate of the critical
temperature of statistical systems. It has frequently been used in ISG
studies to estimate ordering temperatures. However, even in ISGs the
Binder crossing point method is not efficient for determining $T_{\rm c}$
precisely, at least in dimension three, because the $g(T,L)$ curves tend to
lie very close to each other so the estimate of the crossing point is very
sensitive to corrections to scaling.\cite{mari:02}  In the particular case
of the GG, the method is inoperable because the Binder curves do not
intersect, at least for the range of sizes that have been studied here.

For vector systems it is possible to measure a domain-wall-stiffness-like
parameter, the ``current'',\cite{reger:91,reger:93,olson:00,katzgraber:02a} 
which is the rate of change of the free energy with respect to a twist 
angle at the boundaries. For the gauge glass:
\begin{equation}
I(L) = \frac{1}{L} \sum_{i=1}^N \sin(\phi_i - \phi_{i+\hat{x}} -
A_{i\,i+\hat{x}}) \; .
\end{equation}
In this case, the twist is applied along the $\hat{x}$ direction.  As
$[\langle I(L)\rangle ]_{\rm av} = 0$, we actually calculate the
root-mean-square current $I_{\rm rms}$ between two replicas:
\begin{equation}
I_{\rm rms} = \sqrt{[\langle I(L)_\alpha \rangle \langle I(L)_\beta
\rangle]_{\rm av}} \; .
\end{equation}
The root-mean-square currents for different $L$ cross at $T_{\rm c}$ and
splay out at lower temperatures.  They have a finite-size scaling form for
a finite-temperature transition\cite{katzgraber:02a}
\begin{equation}
I = \tilde{I}[L^{1/\nu}(T-T_{\rm c})] \; .
\label{current_scale}
\end{equation}
The intersection of the $I_{\rm rms}(L)$ curves gives a clear indication of 
the value of $T_{\rm c}$. Corrections to finite-size scaling must be allowed
for, but in practice these are weak. The main drawback of this method is
the fact that in current measurements intrinsic sample-to-sample
fluctuations are very strong so that data must be taken on a large number
of independent disorder realizations in thermal equilibrium.

A further important observable is the two-point correlation length. 
In an infinite sample the correlation function
\begin{equation}
G(r_{ij}) = [\langle {\bf S}_i \cdot {\bf S}_j \rangle^2]_{\rm av}
\end{equation}
is of the form
\begin{equation}
G(r_{ij}) \propto r_{ij}^{-(d-2+\eta)} e^{-r_{ij}/\xi(T)} \; ,
\end{equation}
where $i$ and $j$ represent the position of the magnetic moments. $\xi(T)$
is the correlation length, which diverges as $(T-T_{\rm c})^{-\nu}$. In
some systems $G(r)$ has been directly recorded; when $T_{\rm c}$ is known,
the measurement of $G(r)$ at $T_{\rm c}$ on reasonably large samples
provides a very direct measurement of $\eta$. (Measurements have also been
made as a function of anneal time for temperatures lower than $T_{\rm c}$
in the Edwards Anderson ISG.\cite{berthier:02})  Under periodic boundary
conditions there must clearly be a cutoff for $G(r)$ at $r_{ij} = L/2$,
and even before this cutoff the behavior is modified as $G(r_{ij})$ must
become $r$-independent at the cell boundary so as to be compatible with
the periodic boundary conditions. A size-dependent correlation length
$\xi_L$ can be defined
through\cite{ballesteros:00,lee:03,cooper:82,martin:02}
\begin{equation}
\xi_L = \frac{1}{2\sin(|{\bf k}_{\rm min}|/2)}\left[
\frac{\hat{G}({\bf 0})}{\hat{G}({\bf k}_{\rm min})} - 1\right]^{1/2} \; ,
\label{def_xi}
\end{equation}
where $\hat{G}({\bf 0})$ and $\hat{G}({\bf k}_{\rm min})$ are Fourier
transforms of the spatial correlation function $G(r)$\cite{kim:96}, and ${\bf
k}_{\rm min} = (2\pi/L,0,0)$ is the smallest nonzero wave vector. This
second-moment correlation length is in fact an observable having the
dimension of length, which becomes equal to the correlation length in the
limit $L \gg \xi$. It is referred to as the ``size-dependent correlation
length'' even in the limit of $T=T_{\rm c}$ and below, where the
infinite-size correlation length $\xi$ has diverged. The ratio $\xi_L/L$
at $T_{\rm c}$ should be $L$-independent\cite{katzgraber:03a} as the form
of the whole function $G(r)$ scales appropriately with $L$. Curves for the
ratio of the finite-size correlation length to sample length $\xi_L(T)/L$
for different $L$ cross at $T_{\rm c}$ and then splay out at lower
$T$,\cite{palassini:99c,ballesteros:00,lee:03,katzgraber:03a} i.e.,
\begin{equation}
\xi_L/L = {\tilde X}[L^{1/\nu}(T - T_{\rm c} )] \; .
\label{xiscale}
\end{equation}

This makes this observable very attractive for estimating $T_{\rm c}$.
Unfortunately at small and moderate $L$, the ratio $\xi_L/L$ can be very
susceptible to corrections to scaling. In addition, the definition of
$\xi_L$ through Eq.~(\ref{def_xi}) is a convention and other definitions
having the same limiting form at infinite $L$ are equally
plausible.\cite{salas:00} At small and moderate $L$ there are correction
terms $\sim L^{-2}$ from the subleading term in the sine factor in the
conventional definition. There may be a ``lattice artifact'' correction
term\cite{salas:00} of order $1/L$, in addition to the ``true'' correction
term in $L^{-\omega}$ arising from the leading irrelevant operator.
Physically, $\hat{G}(0)$ is the sum over the spins within a box $\sim
L^{d}$ correlated to a central spin and so is equal to the equilibrium
$\chi(L)$. $\hat{G}(2\pi/L)$ contains a positive term from spins close to
the central spin together with a term from some of the spins further from
the central spin which is negative because of the cosine factor. A slight
change in the form of the function $G(r)$ with $L$ will have little effect
on the $L$-dependence of $\hat{G}(0)$ while it can modify the Fourier
transform $\hat{G}(2\pi/L)$ much more drastically.  Hence one can expect
that at low and moderate $L$, $\xi_L/L$ will be much more sensitive to
corrections to finite-size scaling than $\chi(L)$ will. Even for the
canonical ferromagnetic Ising model in two dimensions, the corrections in
$\xi_L/L$ (or to the variant second moment correlation length
$\xi^\prime_L/L$ with a slightly different definition \cite{salas:00}) at
criticality are strong for $L \le 20$, with different correction terms
contributing.\cite{salas:00}  As a consequence, correlation-length data
should as a general rule be treated with caution unless detailed results
exist over a wide range of $L$ from which the various correction
contributions can be evaluated.

Taken over the entire temperature range, ratios such as
$\xi_{2L}(T)/\xi_{L}(T)$ or $\chi(2L,T)/\chi(L,T)$ should be universal
scaling functions of $\xi_{L}(T)/L$.\cite{palassini:99c} If there are
corrections to scaling, scaling curves for different $L$ will not be
identical; this provides an operational method for checking for
corrections to scaling in $\xi_L/L$ not only at or near $T_{\rm c}$, but
over the whole range of temperatures.

Finally, we also record the equilibrium energy per spin. As is well known,
the energy of glassy systems varies smoothly through the ordering
temperature, but the size dependence of the energy has a behavior linked
to the ordering temperature. This is discussed
elsewhere.\cite{katzgraber:03e}

\section{Out of equilibrium methods}
\label{offequil}

Complementary out-of-equilibrium measurements are also carried out. Large
samples of size $L$ are initialized in a disordered (infinite temperature)
state; they are then held at a bath temperature $T$ for an anneal time
$t_{\rm w}$. The spin-glass susceptibility $\chi(t,t_{\rm w})$ is
monitored as a function of $t$. At $T_{\rm c}$ dynamic scaling rules hold;
the spin-glass susceptibility increases proportional to $t^{(2-\eta/z)}$,
as clusters of correlated spins build up over time.\cite{huse:89} This
equation is not valid at very short time scales as $\chi$ is identically
equal to $1$ for a sample of any size and dimension in the totally
disordered state, so the scaling equation will only set in after a
crossover from this initial zero-time value. Even after this crossover,
there can be corrections to scaling in the nonequilibrium susceptibility
data, when only small clusters have been built up. Here the corrections
appear as corrections to {\em finite-time} scaling but are physically
equivalent to corrections to {\em finite-size} scaling. Including the
correction term,
\begin{equation}
\chi(t) = A t^{(2-\eta/z)}(1+Bt^{-w/z} + \cdots) \; ,
\end{equation}
where $w$ is a dynamic correction to scaling exponent analogous to
$\omega$.\cite{parisi:99} In principle $w$ and $\omega$ have no
fundamental reason to be exactly equal but can be expected to be similar.
Finally, if the measurement is continued long enough, the long-time
susceptibility will cross over to a saturation value which is just the
equilibrium susceptibility $\chi(L,T)$ for the measurement temperature $T$
and the sample size $L$ used.

It is clear from a comparison of the equations for the equilibrium
susceptibility and the dynamic susceptibility at $T_{\rm c}$ that for all
intermediate times between the very short time limit and the $L$-dependent
long-time limit, the measurement after a time $t_{\rm w}$ on a large
sample should be equivalent to the measurement at equilibrium on a sample
of size
\begin{equation}
L^{*}= At^{1/z} \; ,
\end{equation}
with $A$ a constant and $z$ the dynamic scaling exponent. $\chi(L)$ and
$\chi(L^{*})$ can be displayed on the same graph; with a judicious choice
of parameters $A(T)$ and $z(T)$ the set of $\chi(L^{*})$ scale onto the
set of equilibrium data $\chi(L)$. This provides us with a direct method
for estimating $z(T)$.  Alternative techniques which have been used for
estimating the value of $z(T)$ independently of the other exponents are
the monitoring of the time variation of the Binder parameter for different
sample sizes\cite{blundell:92} or the scaling of the time-dependent
spin-glass susceptibility $\chi(L,t)$ for different sizes
$L$\cite{olson:00}. The present method is more convenient than the Binder
parameter measurements as the latter are intrinsically noisy.

We have carried out dynamical measurements over a range of temperatures
and not just at the putative ordering temperature. If we assume that there
is an effective dynamic scaling exponent $z(T)$ at each temperature $T$
and not only at $T_{\rm c}$, then at each $T$ we should be able to
translate $\chi(t)$ data into $\chi(L^*)$ data by a suitable choice of
$A(T)$ and $z(T)$. If dynamic scaling continues to hold at other $T$, then
once again it should be possible to make the $\chi(L^*)$ data scale onto
the $\chi(L)$ data. The assumption of an effective $z(T)$ has been made
before in a different context, but applied only at temperatures below
$T_{\rm c}$ (see, for instance. Refs.~\onlinecite{berthier:02} and
\onlinecite{marinari:00}).

Finally, after a long anneal time $t_{\rm w}$ the ultimate configuration 
$\{{\bf S}_i(t_{\rm w})\}$ is registered. The updating procedure is then 
pursued for a further time $t$ considerably less than $t_{\rm w}$ and the 
autocorrelation function decay
\begin{equation}
q(t,t_{\rm w}) = \frac{1}{N}
\sum_{i=1}^N\langle {\bf S}_i(t_{\rm w}+t)\cdot{\bf S}_i(t_{\rm w})\rangle
\label{qt_eq}
\end{equation}
is monitored starting from what can now be considered a quasi-equilibrium
configuration.  In Eq.~(\ref{qt_eq}) ${\bf S}_i$ represents the vector
spins in the plane. In the context of Ising spin glasses it has been
shown\cite{rieger:93,berthier:02,ozeki:01} that the initial decay behavior
of $q(t,t_{\rm w})$ is not dependent on the sample having achieved perfect
thermal equilibrium; as long as the time scale over which $q(t,t_{\rm w})$
is monitored is much shorter than the annealing time $t_{\rm w}$, then the 
form of $q(t,t_{\rm w})$ is characteristic of the infinite-size limit initial
relaxation in equilibrium. Again at $T_{\rm c}$, for a large sample in
equilibrium, the autocorrelation function $q(t,t_{\rm w})$ decays
algebraically as $q(t,t_{\rm w}) \sim t^{-x}$ with\cite{ogielski:85}
\begin{equation}
x = (d - 2 + \eta)/2z \;\;\;\;\;\;\;\; (T = T_{\rm c}) \; .
\end{equation}
Measurements in Ising spin
glasses\cite{rieger:93,kisker:96,ozeki:01,berthier:02} have shown that at
all temperatures below $T_{\rm c}$ the $q(t,t_{\rm w})$ data can be
accurately fitted by this power-law form of decay with a
temperature-dependent effective exponent $x(T)$. For temperatures higher
than $T_{\rm c}$, the decay of $q(t,t_{\rm w})$ has the same initial
power-law form which is now multiplied by a further factor
$\tilde{Q}[t/\tau(T)]$, where $\tilde{Q}$ is a scaling function and
$\tau(T)$ is a relaxation time diverging as $(T-T_{\rm c})^{-z\nu}$ when
$T_{\rm c}$ is approached from above.\cite{ogielski:85} In ISG
measurements $q(t,t_{\rm w})$ shows very little short-time corrections to
scaling; at $T = T_{\rm c}$ and below, $q(t,t_{\rm w})$ for large 
well-annealed samples follows a strict power-law decay beyond very few Monte
Carlo steps after ``zero'' time.

Nonequilibrium measurements can be used in combination with the
equilibrium susceptibility measurements to obtain an estimate of $T_{\rm
c}$ from consistency arguments.\cite{bernardi:96} The effective dynamical
exponent $z(T)$ is measured from the comparison of equilibrium and
nonequilibrium susceptibility measurements. The effective exponent
$\eta(T)$ is measured from the size-dependent equilibrium susceptibility
measurements through $\chi(L,T)/L^2 = L^{-\eta(T)}$. Then, at $T_{\rm c}$,
the relaxation decay exponent $x$ must be equal to $(d - 2 + \eta)/2z$.  
The consistency between the directly measured $x(T)$ and the other
exponents gives an efficient criterion for determining $T_{\rm c}$. This
technique has a number of advantages. $x(T)$ and $z(T)$ can be measured on
almost arbitrarily large samples and so can be rendered free of
corrections to scaling. In addition, they can be measured without the need
to achieve complete thermal equilibrium. The most laborious measurement is
the size-dependence of $\chi(T)$ in equilibrium; however, $\chi(L,T)$
measurements are generally only subject to weak corrections to finite-size
scaling so a limited range of $L$ can be adequate.

The parameters of the off-equilibrium simulations using simple Monte Carlo
updates are listed in Table \ref{simparams_offeq}.

\begin{table}
\caption{
Parameters of the off-equilibrium simulations as a function of space
dimension $d$. $L$ is the size of the system used, $t$ the equilibration
time, and $t_{\rm w}$ is the ``waiting time'' to calculate $q(t,t_{\rm
w})$, Eq.~(\ref{qt_eq}). $N_{\mathrm{samp}}$ is the number of samples used
for the disorder average.
\label{simparams_offeq}
}
\begin{tabular*}{\columnwidth}{@{\extracolsep{\fill}} c c c c c }
\hline
\hline
$d$ & $L$ & $N_{\mathrm{samp}}$ & $t$ & $t_{\rm w}$ \\
\hline
$2$ & $64$ & $500$ & $1.638 \times 10^4$ & $4.069 \times 10^3$ \\
$3$ & $16$ & $200$ & $8.192 \times 10^3$ & $8.192 \times 10^3$ \\
$4$ & $10$ & $200$ & $1.310 \times 10^5$ & $4.096 \times 10^3$ \\

\hline
\hline
\end{tabular*}
\end{table}

\section{Four Dimensions}
\label{4dgg}

In this section we compare different observables as well as critical
exponents and transition temperatures derived from finite-size scaling
arguments for the four-dimensional gauge glass.

\subsection{Root-mean-square current}
\label{4dgg_c}

\begin{figure}
\centerline{\epsfxsize=\columnwidth \epsfbox{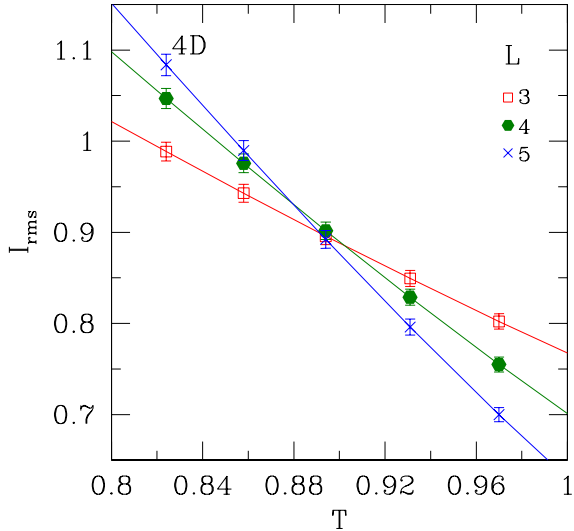}}
\vspace{-1.0cm}
\caption{(Color online)
Data for the the root-mean-square current $I_{\rm rms}$ in
four dimensions. The data show a crossing at $T_{\rm c} \approx 0.89$.
}
\label{4d-current_fig}
\end{figure}

\begin{figure}
\centerline{\epsfxsize=\columnwidth \epsfbox{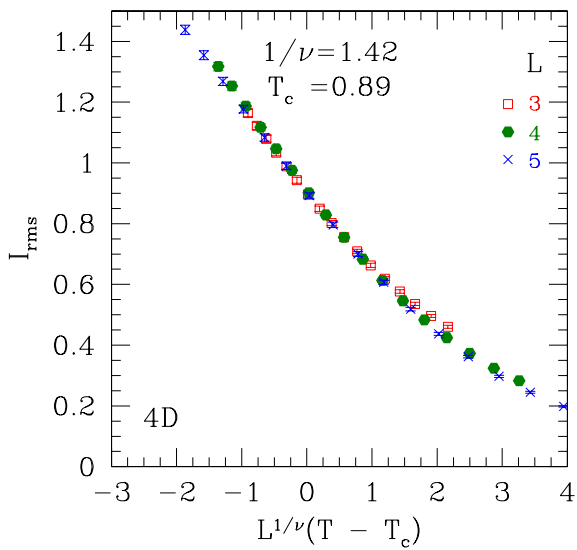}}
\vspace{-1.0cm}
\caption{(Color online)
Scaling of the root-mean-square current $I_{\rm rms}$ in four dimensions
according to Eq.~(\ref{current_scale}). The data scale well for the range
of sizes shown and we estimate $1/\nu \approx 1.42$ with $T_{\rm c}
\approx 0.89$.
}
\label{4d-current_scale_fig}
\end{figure}

Current measurements are made for sizes $L = 3$ -- $5$. The details of the
simulation are shown in Table \ref{simparams_4d}.  The intersection of the
$I_{\rm rms}$ curves shown in Fig.~\ref{4d-current_fig} gives an estimate
$T_{\rm c} = 0.890 \pm 0.015$. This is significantly lower than the value
of $0.96 \pm 0.01$ reported by Reger and Young,\cite{reger:93} although
their data are almost compatible with the present intersection point
within the statistical error bars. Corrections to finite-size scaling
appear to be minimal for these measurements. However, current measurements
are intrinsically noisy with strong sample-to-sample variations at
equilibrium, so it is imperative to average over large numbers of samples.
In Fig.~\ref{4d-current_scale_fig} we show a scaling plot of the data in
Fig.~\ref{4d-current_fig} according to Eq.~(\ref{current_scale}). The data
scale well for the (modest) range of sizes and we estimate $1/\nu = 1.42
\pm 0.03$ together with the above-mentioned estimate of $T_{\rm c}$. The
previously quoted error is estimated by varying the scaling parameters
until the data do not collapse well. This method is also used in all
subsequent estimates of error bars of scaling exponents and critical
temperatures derived from scaling plots.

\subsection{Equilibrium susceptibility}
\label{4dgg_s}

\begin{figure}
\centerline{\epsfxsize=\columnwidth \epsfbox{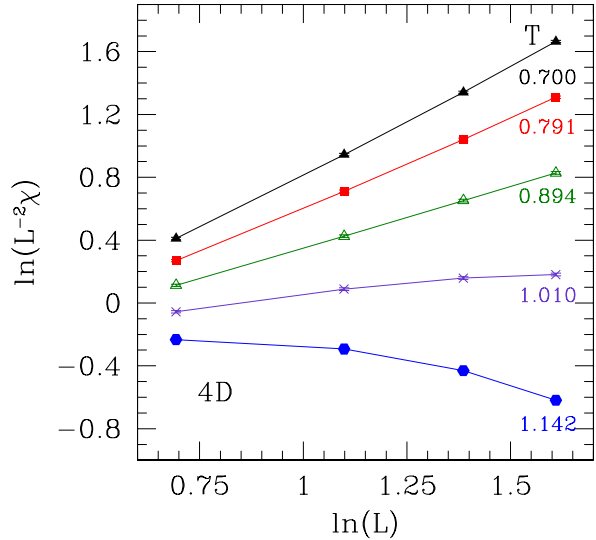}}
\vspace{-1.0cm}
\caption{(Color online)
Data for $\ln(\chi/L^2)$ vs $\ln(L)$ for different temperatures in four
dimensions. At $T_{\rm c}$ one expects $\chi \sim L^{2-\eta}$. The data
show a change in curvature while scanning through $T_{\rm c} \approx
0.89$.
}
\label{4d-logchi_fig}
\end{figure}

The susceptibility data at equilibrium are obtained from the same data set
in Sec.~\ref{4dgg_c} and shown in Fig.~\ref{4d-logchi_fig}. The data are
plotted in the form of curves for $\ln[\chi(L)/L^{2}]$ against $\ln L$ at
different temperatures.  As discussed above, if corrections to scaling are
negligible, this form of plot gives a straight line of slope $-\eta$ at
$T_{\rm c}$, together with curves which bend downward for $T > T_{\rm c}$
and upward for $T < T_{\rm c}$.  For the four-dimensional GG, corrections
are very weak because even including values for $L=2$ the log-log plot of
the data at the temperature closest to $T_{\rm c}$ as estimated above
follows the straight line behavior very closely.

\begin{figure}
\centerline{\epsfxsize=\columnwidth \epsfbox{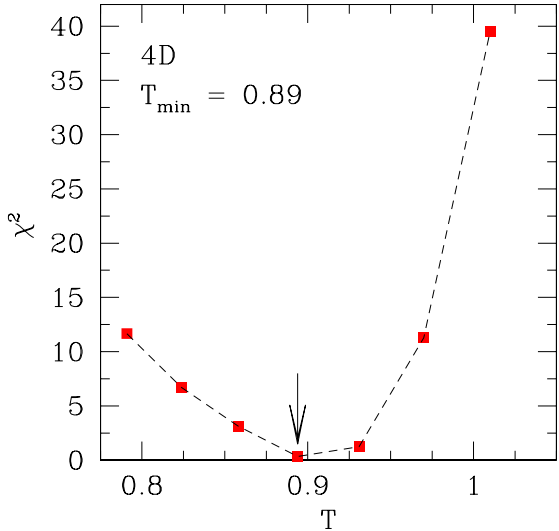}}
\vspace{-1.0cm}
\caption{(Color online)
$\chi^2$-deviation from a straight-line fit for the susceptibility data in
four dimensions shown in Fig.~\ref{4d-logchi_fig}.  The data show a
minimum at $T \approx 0.89$, indicating that the optimal fit happens
around $T = T _{\rm c}$ (marked by an arrow).
}
\label{4d-chi2-sL_fig}
\end{figure}

\begin{figure}
\centerline{\epsfxsize=\columnwidth \epsfbox{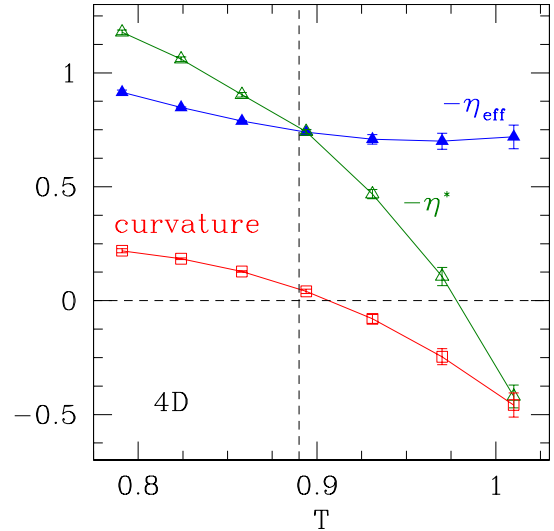}}
\vspace{-1.0cm}
\caption{(Color online)
Data for the effective exponent $\eta_{\rm eff}(T)$ as a function of
temperature in four dimensions. The vertical dashed line marks our
estimate of $T_{\rm c}$ from current and susceptibility measurements,
$T_{\rm c} = 0.89$. In addition, $\eta^*$ is shown, which is an effective exponent
derived from off-equilibrium calculations (described in detail in
Sec.~\ref{offequil}). One expects $\eta_{\rm eff}$ and $\eta^*$ to cross
at $T_{\rm c}$, which is the case in our data within error bars.  Data for
the curvature of the susceptibility shown in Fig.~\ref{4d-logchi_fig} from
a second-order polynomial fit are also displayed. The data cross zero
curvature (horizontal dashed line) at $T \approx 0.90$, a value slightly 
higher than the other estimates.
}
\label{4d-eta_fig}
\end{figure}

In order to obtain an independent estimate for $T_{\rm c}$, we have
plotted the $\chi^2$-deviation from a straight line fit to the data over a
range of temperatures around $T=0.89$, together with the curvature for
three-parameter fits, Figs.~\ref{4d-chi2-sL_fig} and \ref{4d-eta_fig}. The
results indicate straight-line behavior for $T=0.895 \pm 0.015$ which is
an independent measurement of $T_{\rm c}$. In Fig.~\ref{4d-eta_fig} we
also show the effective slope $-\eta(T)$ from the three-parameter fit. The
slope of the straight line at $T_{\rm c}$ gives an estimate of the
critical exponent $\eta= -0.74 \pm 0.03$. As far as we are aware of, this
is the first published estimate for this parameter for the GG in four
dimensions.

\begin{figure}
\centerline{\epsfxsize=\columnwidth \epsfbox{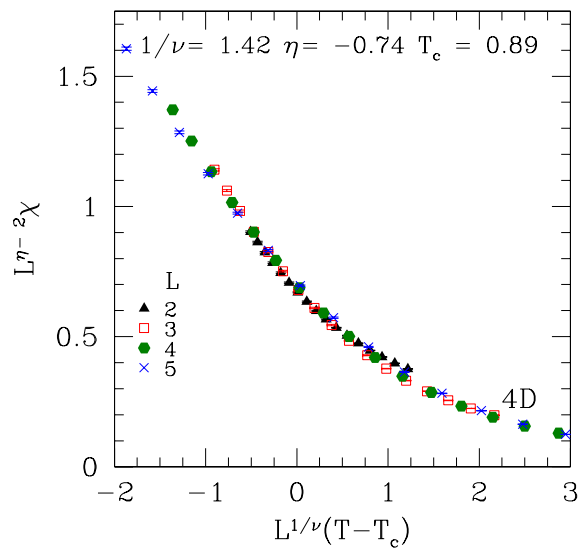}}
\vspace{-1.0cm}
\caption{(Color online)
Scaling plot according to Eq.~(\ref{chi_eqn}) of the susceptibility data
presented in Fig.~\ref{4d-logchi_fig} in four dimensions. The data scale
well for $T_{\rm c} \approx 0.89$, $1/\nu \approx 1.42$, and $\eta \approx
-0.74$. Note that only the data for $L = 2$ show corrections to scaling.
}
\label{4d-susc_scale_fig}
\end{figure}

With values for $T_{\rm c}$ and $\eta$ in hand, the whole $\chi(L,T)$ data
set can be plotted in a standard manner: $\chi(L,T)/L^{2-\eta}$ as a
function of $L^{1/\nu}(T-T_{\rm c})$, adjusting $\nu$ to obtain optimal
scaling. The results lead to $1/\nu=1.42 \pm 0.03$, as shown in
Fig.~\ref{4d-susc_scale_fig}, and are in agreement with an estimate from
the scaling of the currents presented in Fig.~\ref{4d-current_scale_fig}.

\subsection{Correlation length}
\label{4dgg_xi}

\begin{figure}
\centerline{\epsfxsize=\columnwidth \epsfbox{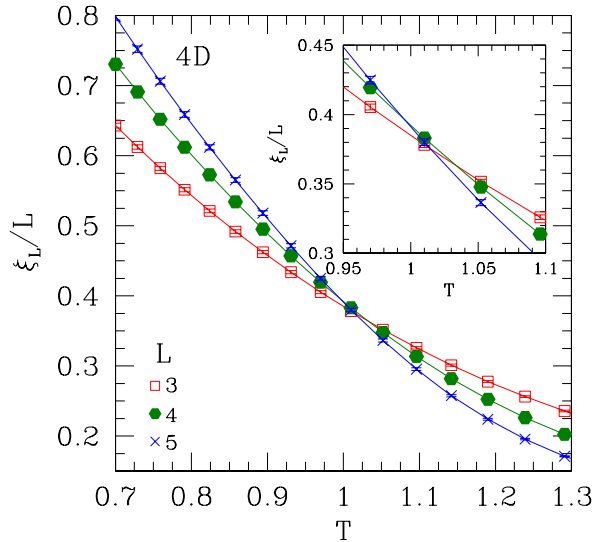}}
\vspace{-1.0cm}
\caption{(Color online)
Correlation length $\xi_L$ divided by $L$ for different system sizes in
four dimensions. The data cross at $T \approx 1.0$ indicating that there
are strong corrections to scaling. The inset shows a zoom of the data in
the main panel focusing on the crossing point around $T \approx 1$.
}
\label{4d-xi_L_fig}
\end{figure}

The data for the ratio $\xi_L(T)/L$ for different sizes are shown in
Fig.~\ref{4d-xi_L_fig}. Although the data appear to cleanly intersect at
one point, it can be seen that the intersection temperatures for these
sizes are around $T \approx 1.0$, and are only approaching the true
ordering temperature, $T_{\rm c} = 0.89$, very slowly with increasing $L$.
The comparison with the behavior observed for $\chi(L)$ or the current is
striking. For each of these two parameters the critical behavior at
$T_{\rm c}$ is virtually correction-free, while for the same range of
small values of $L$, the $\xi_L/L$ intersections give a very poor
indication of the true ordering temperature. There are clearly strong
corrections to scaling at small $L$ which appear to be intrinsic to this
form of measurement.

\subsection{Nonequilibrium susceptibility}
\label{4dgg_non_eq_susc}

\begin{figure}
\centerline{\epsfxsize=\columnwidth \epsfbox{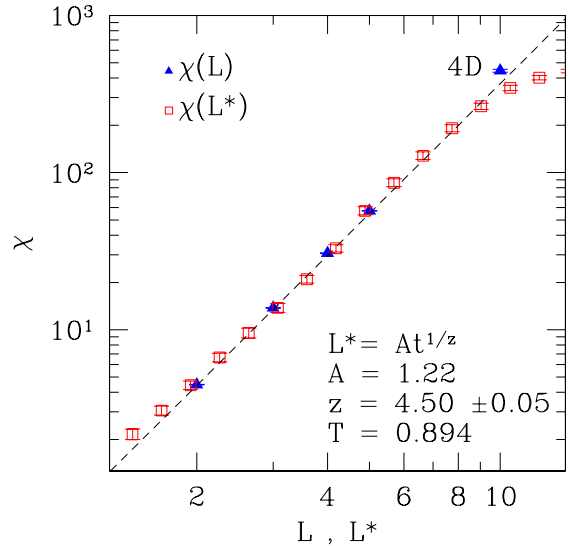}}
\vspace{-1.0cm}
\caption{(Color online)
Spin-glass susceptibility from equilibrium measurements $\chi(L)$ at $T =
0.894$ in four dimensions. $\chi(L^*)$ is the susceptibility determined
from the off-equilibrium simulations with $L^* = At^{1/z}$, $A = 1.22$,
and $z \approx 4.5$. By suitably choosing $L^*$ the data for $\chi(L)$ and
$\chi(L^*)$ fall on a straight line. This allows us to determine the
dynamical critical effective exponent $z(T)$ as a function of temperature.
When $L^{*}(t)$ approaches the sample size (here $L=10$), $\chi(L^{*})$
necessarily saturates. The dashed line is a guide to the eye.
}
\label{4d-chi_LT_fig}
\end{figure}

\begin{figure}
\centerline{\epsfxsize=\columnwidth \epsfbox{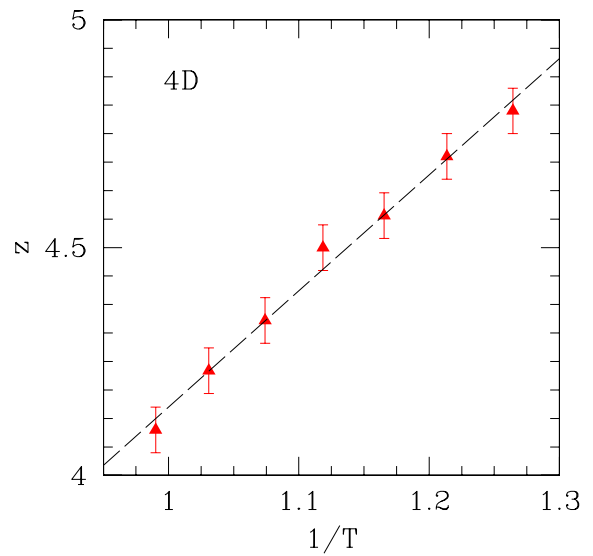}}
\vspace{-1.0cm}
\caption{(Color online)
Dynamical critical exponent $z$ as a function of $1/T$ in four dimensions
as determined from the procedure described in the text and in the caption
of Fig.~\ref{4d-chi_LT_fig}. The data are consistent with $z(T = T_{\rm
c}) = 4.50 \pm 0.05$. Note that the error is estimated ``by eye'': the
data are varied until the two expressions for the susceptibility differ
noticeably thus allowing us to give an upper bound for the error bars. The
dashed line is a guide to the eye.
}
\label{4d-z_fig}
\end{figure}

Large samples ($L=10$ in four dimensions) are initialized in random
(infinite temperature) configurations. They are then put in contact with a
heat bath at a temperature $T$, and the spin-glass susceptibility is
recorded as a function of annealing time. At $T_{\rm c}$ the spin-glass
susceptibility will increase as $t^{(2-\eta)/z}$ until the susceptibility
arrives at the equilibrium value for that size $L$. We present the data in
terms of the scaled time-dependent effective length $L^{*}=At^{1/z}$ at
each temperature. When $A$ and $z$ are suitably chosen, the equilibrium
susceptibility $\chi(L)$ and the nonequilibrium susceptibility
$\chi(L^{*})$ before saturation scale well together. An example for a
temperature close to $T_{\rm c}$, $T = 0.894$, is shown in
Fig.~\ref{4d-chi_LT_fig}. From the data, the dynamical scaling exponent
$z(T)$ can be determined accurately, see Fig.~\ref{4d-z_fig}. The behavior
of the effective dynamical exponent $z(T)$ shows no apparent special
behavior at the critical temperature. Effective temperature-dependent
values $z(T)$ have been reported in ISGs for $T < T_{\rm c}$ (see, for
instance, Ref.~\onlinecite{marinari:00}).  We can conclude from the present
data for the GG that $z(T)$ is a well-defined parameter for a whole range
of temperatures including $T > T_{\rm c}$.  At $T=T_{\rm c}$ we estimate
$z=4.50 \pm 0.05$.

\subsection{Autocorrelation function decay}
\label{4dgg_auto}

\begin{figure}
\centerline{\epsfxsize=\columnwidth \epsfbox{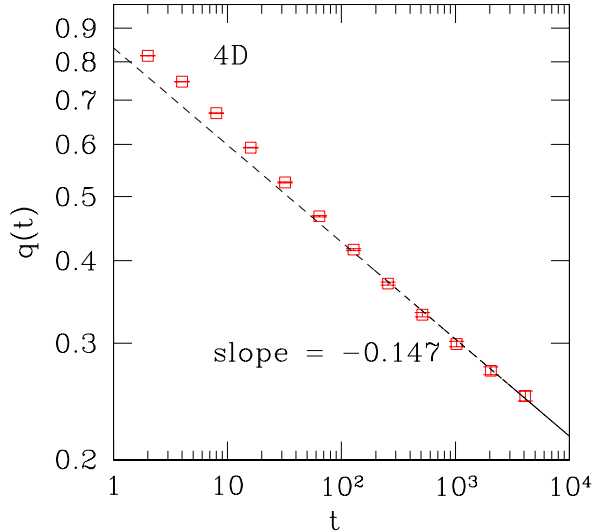}}
\vspace{-1.0cm}
\caption{(Color online)
Data for the autocorrelation function $q(t)$ in four dimensions as a
function of time (measured in Monte Carlo steps). The dashed line has
slope $-x = -0.147$ from which we can determine $\eta^*$, a ``dynamical''
effective exponent $\eta$. Details are described in the text.
}
\label{4d-q_fig}
\end{figure}

In the GG we have found unexpectedly that $q(t)$ only assumes a pure
power-law behavior after a relatively long time, of the order of 100 
Monte Carlo steps, hereafter referred to as MCS,
(as compared with $\sim 5$ MCS in ISGs), see Fig.~\ref{4d-q_fig}.  For
earlier times the decay is affected by short-time corrections.  In
practice this means that the measurements of $x(T)$ as defined above are
less precise than in ISGs as they are limited at short times by the
correction term and at long times by the condition that the maximum
$t_{\rm w}$ should be much less than $t$.

We compare, for a set of temperatures around $T_{\rm c}$, the value of
$\eta(T)$ directly obtained from the equilibrium susceptibility
measurements with a value calculated indirectly from the $x(T)$ and $z(T)$
data, $\eta^{*}(T)=(d-2)-2z(T)x(T)$. We expect $\eta(T)=\eta^{*}(T)$ at
$T=T_{\rm c}$. This can be seen in Fig.~\ref{4d-eta_fig} where the
vertical dashed line marks our estimate of $T_{\rm c}$. The two curves
cross at $T=0.90 \pm 0.01$, $\eta = - 0.74 \pm 0.02$, which again defines
the ordering temperature and the critical exponent $\eta$.  As for this
particular system $T_{\rm c}$ is already reliably established from the
previously mentioned methods, the comparison provides a critical benchmark
test for the nonequilibrium technique. The good agreement with the
standard methods, such as susceptibility scaling and the crossing of 
root-mean-square currents in this case shows that this method is reliable, 
implying that it can be used for other systems.

\section{Three Dimensions}
\label{3dgg}

A number of estimates have already been given of the ordering temperature
and critical exponents of the GG in dimension three.\cite{olson:00} In
this section we present estimates for the critical temperature and
exponents as derived from our simulations. In comparison to the
four-dimensional data presented in Sec.~\ref{4dgg} we find strong
corrections to scaling in three dimensions. In what follows the different
results for different observables are presented.

\subsection{Root-mean-square current}
\label{3dgg_c}

\begin{figure}
\centerline{\epsfxsize=\columnwidth \epsfbox{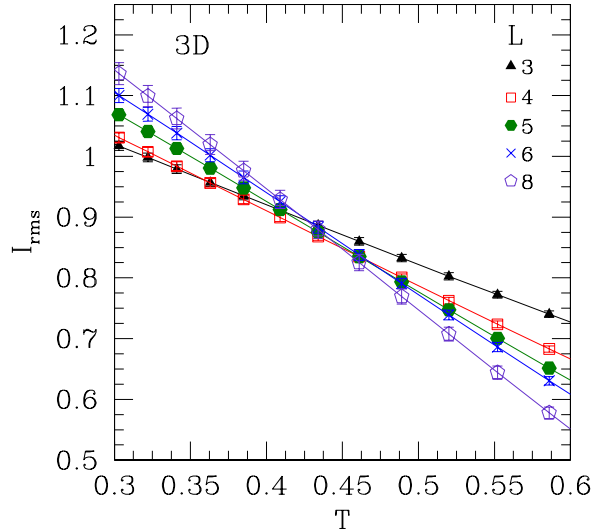}}
\vspace{-1.0cm}
\caption{(Color online)
Root-mean-square current $I_{\rm rms}$ as a function of temperature for
the three-dimensional GG. The data show a crossing at $T \approx 0.46$.
Note that the data for $L = 3$ show strong corrections to scaling.
}
\label{3d-current_fig}
\end{figure}

\begin{figure}
\centerline{\epsfxsize=\columnwidth \epsfbox{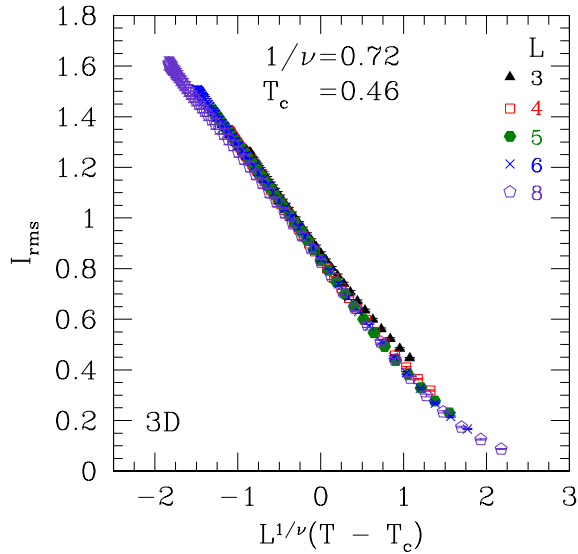}}
\vspace{-1.0cm}
\caption{(Color online)
Scaling of the root-mean-square current $I_{\rm rms}$ in three dimensions
according to Eq.~(\ref{current_scale}). We see acceptable scaling of the
data around $T = 0.46$. Deviations at higher $T$ are presumably due to
corrections to scaling. This plot is for $1/\nu = 0.72$ and $T_{\rm c} =
0.46$. 
}
\label{3d-current_scale_fig}
\end{figure}

Olson and Young\cite{olson:00} estimated that $T_{\rm c} = 0.47 \pm 0.03$
from the intersection of the root-mean-square current induced by an
infinitesimal twist along the boundaries. The present current data are
shown in Fig.~\ref{3d-current_fig}.  Our results agree with the data of
Ref.~\onlinecite{olson:00} within error bars. A consistent intersection
point for all the data taken together occurs at $T=0.46 \pm 0.01$, with
only the data for $L = 3$ lying slightly below the region of intersection.
We conclude that this temperature indeed represents the correct ordering
temperature for the system, corrections to finite-size scaling being small
for $L \geq 4$. The precision is limited mainly by the statistical error
bars, as the curves for different sizes lie rather close together. A
scaling plot of the root-mean-square current according to
Eq.~(\ref{current_scale}) is shown in Fig.~\ref{3d-current_scale_fig} for
$1/\nu = 0.72$ and $T_{\rm c} = 0.46$.

\subsection{Equilibrium susceptibility}
\label{3dgg_s}

\begin{figure}
\centerline{\epsfxsize=\columnwidth \epsfbox{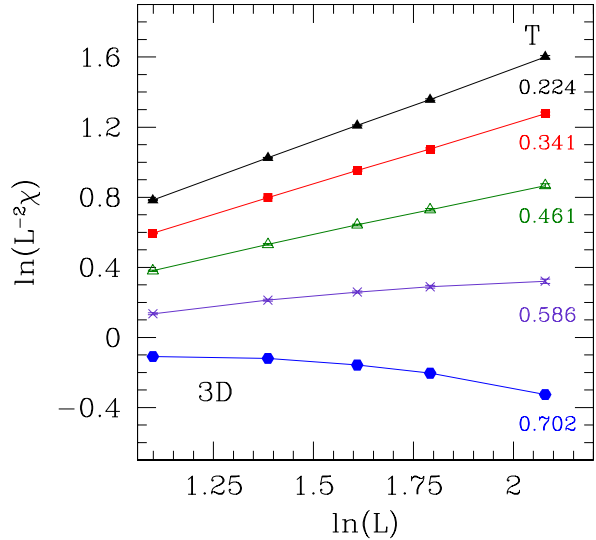}}
\vspace{-1.0cm}
\caption{(Color online)
Data for $\ln(\chi/L^2)$ vs $\ln L$ for different temperatures in three
dimensions. The data show a change in curvature in the log-log plot while
scanning through the $T_{\rm c}$ estimate from the root-mean-square
currents, $0.46$. At $T_{\rm c}$ we expect $\chi \sim L^{2-\eta}$.
}
\label{3d-logchi_fig}
\end{figure}

\begin{figure}
\centerline{\epsfxsize=\columnwidth \epsfbox{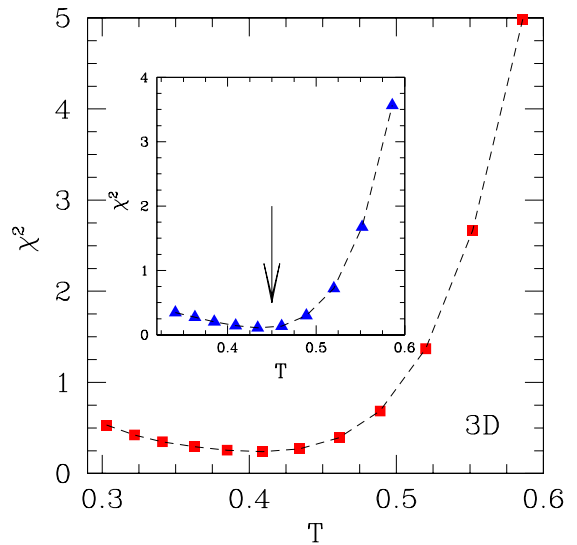}}
\vspace{-1.0cm}
\caption{(Color online)
$\chi^2$-deviation from a straight-line fit for the susceptibility data in
three dimensions. The main panel shows data for a fit with no corrections
to scaling due to lattice artifacts, whereas in the inset we show the data
where corrections to scaling are included in the fits of $\ln(\chi/L^2)$.
The data in the inset show a minimum at $T \approx 0.45$, in agreement
with data from Ref.~\onlinecite{olson:00} and with the root-mean-square
current estimate from Sec.~\ref{3dgg_c}.
}
\label{3d-chi2-sL_fig}
\end{figure}

A log-log plot of the equilibrium susceptibility divided by $L^2$ as
function of $L$ for different temperatures is shown in
Fig.~\ref{3d-logchi_fig}.  In absence of corrections to scaling this plot
should be straight at $T_{\rm c}$. We see that for $T > 0.46$ the data
curve down, indicative that we are above the critical temperature. For $T
< 0.46$ the data show a slight upward curvature,\cite{curve} in agreement 
with $T < T_{\rm c}$. Using Eq.~(\ref{chi_eqn_2}) at $T = T_{\rm c}$ we obtain
$\eta=-0.47 \pm 0.02$ and a correction factor $\sim (1 - AL^{-\omega})$
with $\omega \approx 2.5$ and $A = 0.6$.  For temperatures below $T_{\rm
c}$ a higher effective $\eta(T)$ is obtained at each temperature and the
correction factor appears to be temperature-independent. The correction
term is almost negligible for $L \geq 4$; if we make straight line fits to
$\ln[\chi(T)/L^2]$ against $\ln L$ for $L=4$ to $8$, we find $\chi^2$
values which increase sharply above $T \approx 0.40$, and which rise
slowly for temperatures below $T \approx 0.40$, see Fig.~\ref{3d-chi2-sL_fig}.

\begin{figure}
\centerline{\epsfxsize=\columnwidth \epsfbox{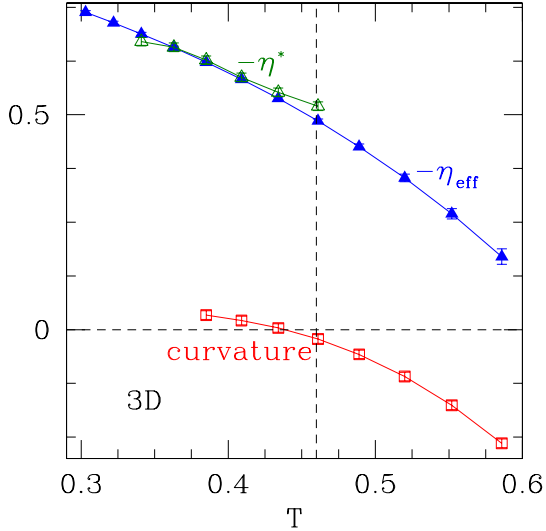}}
\vspace{-1.0cm}
\caption{(Color online)
Data for the effective exponent $\eta_{\rm eff}(T)$ as a function of
temperature in three dimensions. Also plotted is $\eta^*$ as estimated from
off-equilibrium simulations. Both exponents do not cross cleanly making it
difficult to determine $T_{\rm c}$. The vertical dashed line marks our
estimate for $T_{\rm c}$, $0.46$. Also shown is the curvature of the
susceptibility data (see Fig.~\ref{3d-logchi_fig}) from a second-order
polynomial fit. The data cross zero curvature (horizontal dashed line) at $T =
0.44 \pm 0.01$, a value slightly lower than other estimates of $T_{\rm c}$.
}
\label{3d-eta_fig}
\end{figure}

In both Ising spin glasses and GGs the first term in the RG
$\epsilon$-expansion for the leading irrelevant operator is $\omega(d)\sim
(6-d)$\cite{dedominicis:03,bray:03} implying that $\omega$ will be high in
dimension three. High-temperature series expansion data show that $\omega$ is
greater than $3$ in dimension three for the ISG\cite{klein:91}. If we make
the plausible assumption that this is the case also for the GG, the
leading correction term will be due to the lattice artifact so the
correction factor is $[1+ A L^{-(2 - \eta)}]$.  Fits to all 3D GG
$\chi(L,T)$ data from $L=2$ to $8$ using this correction factor give a
clear minimum in $\chi^2$ as a function of temperature at $T=0.45 \pm
0.02$ (inset to Fig.~\ref{3d-chi2-sL_fig}), which we can identify with
$T_{\rm c}$ . The effective exponent $\eta_{\rm eff}(T)$ is shown in
Fig.~\ref{3d-eta_fig}.

\begin{figure}
\centerline{\epsfxsize=\columnwidth \epsfbox{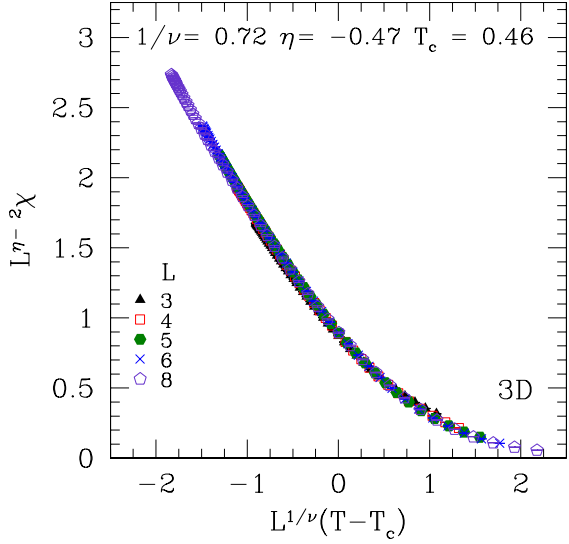}}
\vspace{-1.0cm}
\caption{(Color online)
Scaling plot according to Eq.~(\ref{chi_eqn}) of the susceptibility data
presented in Fig.~\ref{3d-logchi_fig} in three dimensions. Optimal scaling
is obtained for $T_{\rm c} \approx 0.46$, $1/\nu \approx 0.72$, and $\eta
\approx -0.47$.
}
\label{3d-susc_scale_fig}
\end{figure}

An overall scaling plot of $\chi(L,T)/L^{2-\eta}$ as a function of
$L^{1/\nu}(T-T_{\rm c})$ for $L \ge 4$ is shown in
Fig.~\ref{3d-susc_scale_fig}. We estimate $1/\nu = 0.72 \pm 0.02$.

\subsection{Binder parameter}
\label{3dgg_gL}

\begin{figure}
\centerline{\epsfxsize=\columnwidth \epsfbox{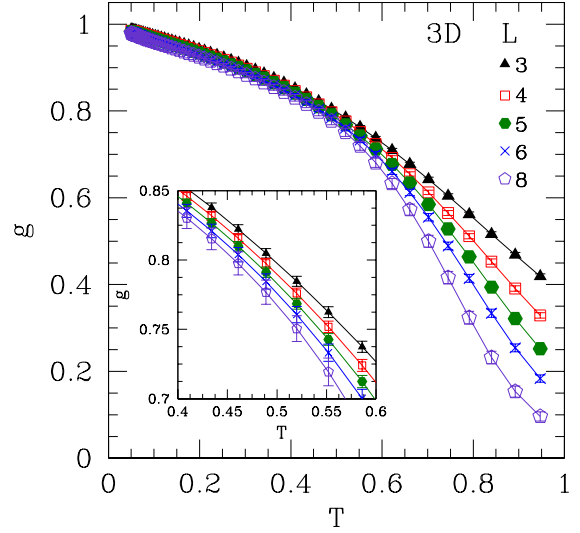}}
\vspace{-1.0cm}
\caption{(Color online)
Data for the Binder ratio $g$ as a function of temperature for several
system sizes. We see that the data do not cross for $T = T_{\rm c} \approx
0.46$ and also do not splay for $T$ smaller than $T_{\rm c}$. The inset
zooms into the region $T \leq 0.6$.
}
\label{3d-binder_fig}
\end{figure}

In spin glasses the Binder parameter $g$ defined in Eq.~(\ref{binder}) is
independent of system size at the transition temperature as it is
proportional to a function which only depends on $L^{1/\nu}(T - T_{\rm
c})$. Consequently different lines at different temperatures cross at
$T_{\rm c}$. In vector systems this is not the case: data for different
system sizes splay for $T > T_{\rm c}$ but not for $T < T_{\rm c}$, as can
be seen in Fig.~\ref{3d-binder_fig}. From the data one can, at best,
estimate $T_{\rm c}$ roughly because the data do not cross. The same
behavior is found in the three-dimensional $XY$ spin
glass.\cite{kawamura:01} This has been ascribed to the ordering being
chiral in this model. As no chirality can be defined for the gauge glass,
this is not the case here and the behavior of $g$ for the GG remains to be
understood.

\subsection{Correlation length}
\label{3dgg_xi}

\begin{figure}
\centerline{\epsfxsize=\columnwidth \epsfbox{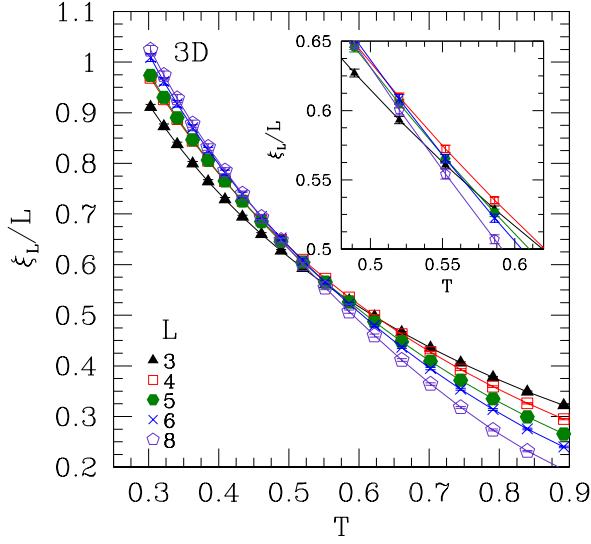}}
\vspace{-1.0cm}
\caption{(Color online)
Correlation length $\xi_L$ divided by $L$ for different system sizes in
three dimensions. The data cross at $T \approx 0.5$ but only splay
slightly making it difficult to give a precise estimate of the transition
temperature.
}
\label{3d-xi_L_fig}
\end{figure}

The data for the ratio $\xi_L(T)/L$ are shown in Fig.~\ref{3d-xi_L_fig}.  
As in four dimensions there are strong finite-size correction effects; in
addition, the curves for different $L$ lie close together.  The
intersections between curves for different $L$ change with temperature in
such a way that from these data alone it would be very hard to identify
$T_{\rm c}$ to better than $T_{\rm c} = 0.50 \pm 0.05$.

\subsection{Nonequilibrium susceptibility}
\label{3dgg_non_eq_susc}

\begin{figure}
\centerline{\epsfxsize=\columnwidth \epsfbox{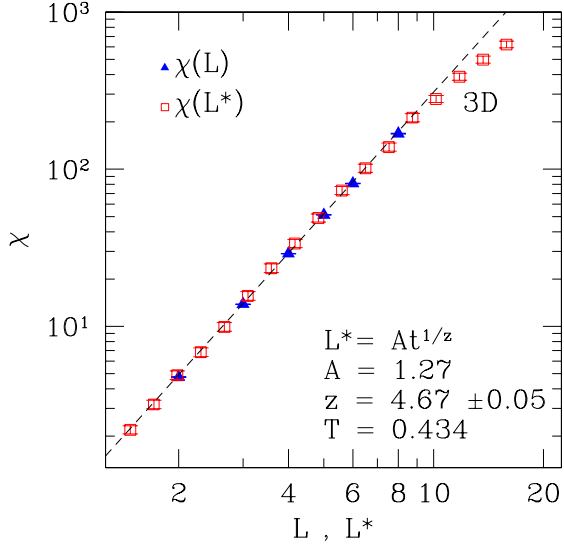}}
\vspace{-1.0cm}
\caption{(Color online)
Spin-glass susceptibility from equilibrium measurements $\chi(L)$ at $T =
0.434$ in three dimensions. $\chi(L^*)$ is the susceptibility determined
from the off-equilibrium simulations with $L^* = At^{1/z}$, $A = 1.27$,
and $z \approx 4.67$. We choose $L^*$ so that the data for $\chi(L)$ and
$\chi(L^*)$ fall on a straight line allowing us to determine the dynamical
critical effective exponent $z(T)$ as a function of temperature. When
$L^{*}(t)$ approaches the sample size (here $L=16$), $\chi(L^{*})$
necessarily saturates. The dashed line is a guide to the eye.
}
\label{3d-chi_LT_fig}
\end{figure}

\begin{figure}
\centerline{\epsfxsize=\columnwidth \epsfbox{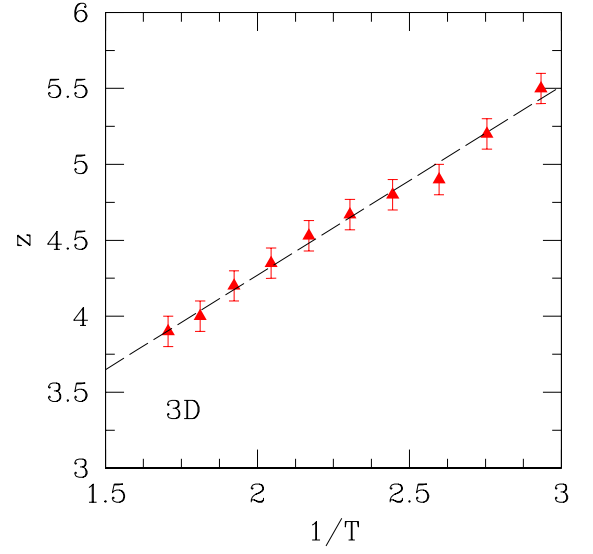}}
\vspace{-1.0cm}
\caption{(Color online)
Dynamical critical exponent $z$ as a function of $1/T$ in three
dimensions. For details see the text. The data are consistent with $z(T =
T_{\rm c}) = 4.7 \pm 0.1$, where the error is estimated by shifting the
data until they visibly do not agree. The dashed line is a guide to the
eye.
}
\label{3d-z_fig}
\end{figure}

In just the same way as in four dimensions the nonequilibrium
susceptibility is recorded for large system sizes, $L=16$.  The effective
dynamical exponent $z(T)$ is estimated from the comparison of the
time-dependent nonequilibrium susceptibility and the size-dependent
equilibrium susceptibility, Fig.~\ref{3d-chi_LT_fig}, and the effective
dynamical critical exponent $z(T)$ is shown as a function of temperature
in Fig.~\ref{3d-z_fig}.  At $T=0.46$ we obtain $z(T)=4.7 \pm 0.1$, in
agreement with the results from Ref.~\onlinecite{olson:00} but with 
smaller error bars.

\subsection{Autocorrelation function decay}
\label{3dgg_auto}

\begin{figure}
\centerline{\epsfxsize=\columnwidth \epsfbox{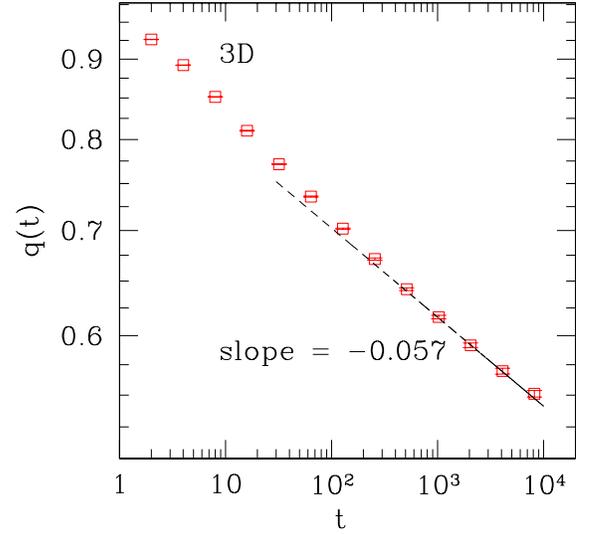}}
\vspace{-1.0cm}
\caption{(Color online)
Data for the autocorrelation function $q(t)$ in three dimensions as a
function of time (measured in Monte Carlo steps). The dashed line is a
guide to the eye to emphasize the asymptotic slope of $q(t)$ in a log-log
plot.
}
\label{3d-q_fig}
\end{figure}

Corrections to finite time scaling in three dimensions extend to times of
the order of 100 MCS, limiting the precision of the measurement of the
exponent $x(T)$, as in four dimensions. For this particular system the
directly measured effective exponent $\eta(T)$ and the indirectly
estimated $\eta^{*}(T)$ are very similar over a range of temperatures, as
shown in Fig.~\ref{3d-eta_fig}. Hence it is not possible to obtain an
independent measurement of $T_{\rm c}$ using the consistency criterion
outlined above (see Fig.~\ref{3d-q_fig}).

\section{Two Dimensions}
\label{2dgg}

The gauge glass in two dimensions has been studied extensively in
Refs.~\onlinecite{katzgraber:02a} and \onlinecite{katzgraber:03a}.  In
particular it was found that for this model $T_{\rm c} = 0$, inferred from
the values of critical exponents and several tests made. In this section
we present results for off-equilibrium simulations, as well as another
test which shows that the system orders at zero temperature in two
dimensions.

\subsection{Correlation length}

\begin{figure}
\centerline{\epsfxsize=\columnwidth \epsfbox{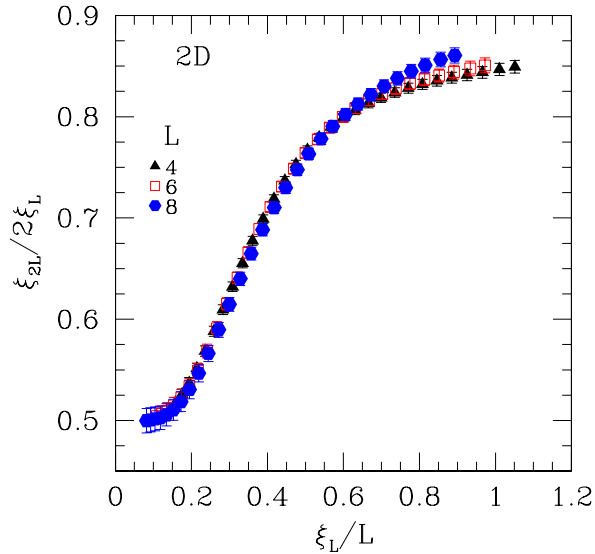}}
\vspace{-1.0cm}
\caption{(Color online)
Scaling plot of the data for $\xi_{2L}/2 \xi_L$ vs $\xi_L/L$ in two
dimensions. The deviations at large $\xi_L/L$ suggest corrections to
scaling. If $T_{\rm c}$ is finite, the curves must go through
$\xi_{2L}/2 \xi_L = 1$ at $T =T_{\rm c}$. The data only tend asymptotically
to $1$ proving that $T > T_{\rm c}$.
}
\label{2d_xi2xi_fig}
\end{figure}

The GG in dimension two has been extensively
studied.\cite{fisher:91,gingras:92,akino:02,reger:93,katzgraber:02a,simkin:96,granato:98}
Measurements on various observables, in particular the correlation length,
have established that the ordering temperature is either zero or much
lower than our lowest measuring temperature,
$T=0.13$.\cite{katzgraber:03a} In Fig.~\ref{2d_xi2xi_fig} we show
correlation length data presented in the form of plots of the ratio
$\xi(2L)/2\xi(L) \equiv \xi_{2L}/2\xi_L$ against $\xi(L)/L$. In the
absence of corrections to scaling all points should fall on a single
scaling curve, which for a finite ordering temperature should pass through
$\xi(2L)/2\xi(L)=1$ at $T=T_{\rm c}$ [here $\xi(L)/L$ is independent of
$L$].\cite{palassini:99c} For ordering which takes place only at zero
temperature the ratio should tend asymptotically to a value which can be 
less than $1$. The data show this form of behavior, demonstrating 
conclusively that for this system $T_{\rm c}=0$.

\subsection{Nonequilibrium scaling} 

\begin{figure}
\centerline{\epsfxsize=\columnwidth \epsfbox{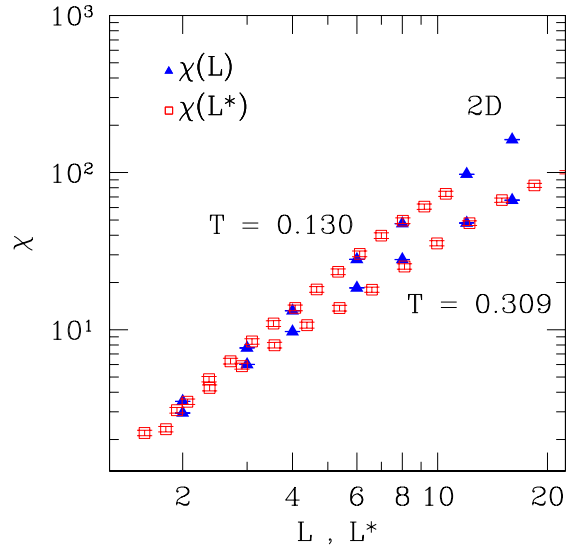}}
\vspace{-1.0cm}
\caption{(Color online)
Examples of excellent dynamic scaling well into the paramagnetic region 
in two dimensions. Spin-glass susceptibility from equilibrium measurements 
$\chi(L)$ (triangles) and $\chi(L^{*})$ from scaled dynamical measurements 
(squares) at two temperatures: $T = 0.130$ (upper set of curves) and 
$T = 0.309$ (lower set of curves). 
}
\label{2d-chi_LT_fig}
\end{figure}

\begin{figure}
\centerline{\epsfxsize=\columnwidth \epsfbox{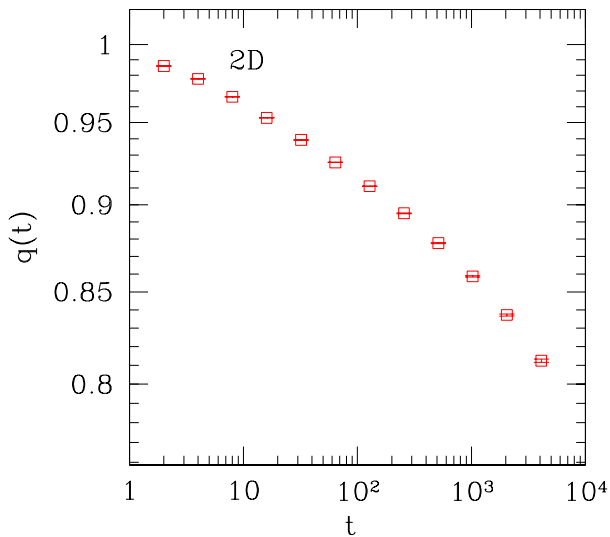}}
\vspace{-1.0cm}
\caption{(Color online)
Data for the autocorrelation function $q(t)$ in two dimensions as a
function of time (measured in Monte Carlo steps). Here $L=64$ and $T=0.13$.
The curvature of the log-log plot demonstrates yet again that any freezing 
is well below this temperature.
}
\label{2d-q_fig}
\end{figure}

Measurements of the equilibrium susceptibility $\chi(L)$ for $L=4$ to $16$
(or to $24$ above $T = 0.20$) and of the nonequilibrium susceptibility
$\chi(t)$ for samples of size $L=64$ are shown in
Fig.~\ref{2d-chi_LT_fig}.  As for the other dimensions, the
nonequilibrium data are scaled using $L^{*}=At^{1/z}$ with $A$ and $z$
chosen at each temperature so that the two sets of susceptibility data
scale together. The scaling is particularly satisfactory because a wide
range of $L$ could be used for the equilibrium measurements, and because
the nonequilibrium sample sizes have been chosen such that for all
measurements $L^{*}$ is much smaller than the sample size, so there are no
saturation effects at long times. This demonstrates once again, but more
clearly than for higher dimensions, that a dynamical exponent $z(T)$ can
be defined relating the annealing time $t$ to an effective length
$L^{*}(t)$ for temperatures well above any critical temperature.  The
scaling holds to within the present statistical accuracy for the whole
range of $L$ from $L=4$ to $L=24$, or alternatively for anneal times from
$2$ MCS to over $16000$ MCS. It appears that the dynamic scaling concept
is ubiquitous and is not just valid at critical points. Finally,
Fig.~\ref{2d-q_fig} shows data for the autocorrelation function $q(t)$.
One can clearly see that the data, presented in a log-log plot, are
strongly curved suggesting that the standard scaling functions for a
finite-temperature transition do not apply.

\section{Comparison of Techniques}
\label{summary_comp}

We can draw comparisons between the different techniques for estimating
$T_{\rm c}$ and the critical exponents. It should be noted that
equilibrium data for the different observables studied are taken within
the same runs which allows a comparison of the relative statistical
precision of the different measurements for exactly the same computational
effort. As an example, for measurements made at $L=8$ at $T=0.461$ in
dimension three and with the simulation parameters listed in Table
\ref{simparams_3d}, the spin-glass susceptibility $\chi$ is accurate to
$0.5\%$, the correlation length $\xi(L)$ to $0.7\%$, the current $I_{\rm
rms}$ to $1.6\%$ (and the internal energy to $0.02\%$)  with the quoted
errors purely statistical.

The major difficulty in the interpretation of the data is not statistics
but corrections to scaling. In the gauge glass the current method turns
out to be relatively insensitive to corrections to finite-size scaling and
is therefore already reliable for relatively small samples. Averaging over
large numbers of samples is, however, essential for these measurements
because of strong intrinsic sample to sample fluctuations. In ISGs
measurements of the critical behavior of the domain-wall stiffness (see,
for instance, Ref.~\onlinecite{hukushima:99}) could well provide an equally
efficient and reliable method to estimate $T_{\rm c}$, applicable quite
generally.

Direct analysis of the spin-glass susceptibility $\chi(L,T)$ on its own is
also reliable and efficient, particularly as the statistical precision of
this parameter is high. At $T = T_{\rm c}$ there is pure power-law
behavior, $\chi(L,T)/L^2 \sim L^{-\eta(T)}$. The onset of deviations from
the pure power-law behavior below and above the ordering temperature
provides a clear signature for $T_{\rm c}$.

In dimension four, even with the restricted range of $L$ available, there
is a downward deviation from the pure power law as soon as $ T > T_{\rm
c}$, and an upward deviation from the pure power law for temperatures $T <
T_{\rm c}$. Corrections appear to be negligible as the straight line fit
at $T_{\rm c}$ is excellent even for $L=2$.

In dimension three the onset of deviation from a pure power law to a
downward curvature at temperatures above $T\sim 0.46$ gives a clear
indication of $T_{\rm c}$ but the low-temperature upturn is weak.
Corrections are present as there is a weak curvature at small $L$ for all
temperatures. If the plausible assumption is made that the leading
correction to scaling is the leading analytic term, a precise estimate for
$T_{\rm c}$ is obtained from fits using all the $\chi(L,T)$ data over the 
whole range of $L$. This estimate is consistent with the value from the 
current measurements.

The susceptibility method (see Secs.~\ref{4dgg_s} and \ref{3dgg_s})  
becomes progressively more precise when data for larger samples are
available. In the present case, data up to $L=5$ only are sufficient for
dimension four, but in dimension three it would have been very helpful to have
had data up to $L=12$; with such data in hand it would be possible to
pinpoint the transition temperature to even higher precision from
susceptibility measurements alone.

The technique involving the consistency of observables deduced from
nonequilibrium scaling and equilibrium susceptibility gives an estimate
for $T_{\rm c}$ which has been found to be in excellent agreement with
those obtained from the other methods for dimension four. This despite the
presence of short-time corrections to scaling in $q(t)$ that were
unexpected.  In dimension three consistency between alternative methods of
estimating parameters holds over a range of temperatures rather than at a
unique temperature which could be identified with $T_{\rm c}$. This
accidental consistency has not been observed in other systems and remains
to be understood, but means that the technique in this particular case
cannot be used to estimate $T_{\rm c}$ with any precision. The
nonequilibrium measurements are relatively economical in computational
resources as full thermodynamic equilibration is not required.

Estimates of $T_{\rm c}$ from the intersections of correlation length
ratios $\xi(L)/L$ turn out to be misleading in dimension four because of
strong intrinsic corrections to finite-size scaling for the small or
moderate system sizes to which computational resources generally limit
measurements in spin-glass systems at high dimensions. This parameter thus
gives a qualitative indication that spin-glass ordering is occurring but
it is not reliable for extracting precise values of the ordering
temperature from small $L$ data even in a situation where other
measurements can give consistent and satisfactory estimates for an
equivalent range of $L$. It can be noted that while corrections to
finite-size scaling in measurements of $\chi(L)$ appear to be negligible
for the GG in dimension four, the corrections for $\xi(L)/L$ are still very
strong at size $L=5$. Even in dimension three where the corrections are
weaker, this observable remains a poor tool for estimating the transition
temperature. Strong correction effects in $\xi(L)/L$ can also be seen in
data reported for $XY$ spin glasses\cite{lee:03} and have been found in
certain ISGs. Therefore the correction effect appears to be generic and
this technique should be applied only with caution.

Finally, estimates from the Binder parameter are inappropriate for the GG
as the curves for different system sizes do not intersect for the range of
$L$ over which measurements are carried out. It is possible that there are
strong finite-size corrections for this parameter so that intersections
would only be seen for much larger $L$.

Once $T_{\rm c}$ has been estimated, the susceptibility measurements at
that temperature give immediately the corresponding value of the critical
exponent $\eta$. The precision for this parameter is limited entirely by
the accuracy with which $T_{\rm c}$ is known. The exponent $\nu$ is
estimated from scaling plots of current data and from scaling plots of
susceptibility data. Once again precision is limited by the accuracy with
which $T_{\rm c}$ has been determined.  The dynamical exponent $z$ can be
measured accurately from a comparison of equilibrium and nonequilibrium
susceptibilities. The data show that $z(T)$ can be estimated operationally
over a wide range of $T$ including temperatures well above the ordering
temperature [$z(T)$ varies continuously through the ordering temperature].

In the light of this extensive analysis, it would seem appropriate to
critically re-assess the estimates for the critical exponents in the ISG
family.

\section{Critical Parameters as a Function of Space Dimension}
\label{gg_exponents}

The critical parameters for the GG obtained from this and earlier work are
listed in Table \ref{crit_table}.

\begin{table}
\caption{
Critical temperature $T_{\rm c}$ and critical exponents for the GG in
different space dimensions $d$ and from different references. In the table
KC represents results from this work, whereas RY represents
Ref.~\onlinecite{reger:93} by Reger and Young, OY
Ref.~\onlinecite{olson:00} by Olson and Young, and K
Ref.~\onlinecite{katzgraber:03a} by Katzgraber.
\label{crit_table}
}
\begin{tabular*}{\columnwidth}{@{\extracolsep{\fill}} c l c c c c c }
\hline
\hline
$d$ & Reference & $T_{\rm c}$ & $\eta$ & $\nu$ & $z$ & $\omega$\\
\hline
$4$ & RY        & $0.96(1)$   &            & $0.70(15)$ &           &          \\
$4$ & KC        & $0.89(1)$   & $-0.74(3)$ & $0.70(1)$  & $4.50(5)$ &          \\
\hline
$3$ & OY        & $0.47(3)$   & $-0.47(7)$ & $1.39(20)$ & $4.2(6)$  &          \\
$3$ & KC        & $0.460(15)$ & $-0.47(2)$ & $1.39(5)$  & $4.7(1)$  & $2.5(5)$ \\
\hline
$2$ & K          & $0$         & $0$        & $2.56(20)$ & $\infty$  &         \\
\hline
\hline
\end{tabular*}
\end{table}

In both dimensions three and four the GG ordering temperatures $T_{\rm c}$ are
roughly half of those for the Gaussian ISG.  By interpolation between the
zero-temperature current stiffness exponents $\theta(d)$ in dimensions 1,
2, and 3,\cite{katzgraber:02a} we can estimate the lower critical dimension
for the GG as the dimension at which $\theta(d)$ passes through
zero: $d_{\rm lcd} \approx 2.5$, very similar to that of the ISG systems
with continuous interaction distributions\cite{bouchaud:02,exponents} (see
Table \ref{crit_table_ISG} and
Refs.~\onlinecite{parisi:96,marinari:98,campbell:00,bhatt:88}).

\begin{table}
\caption{
Critical temperature $T_{\rm c}$ and critical exponents for the ISG in
different space dimensions $d$ with Gaussian interactions. In the table
PRR represents Ref.~\onlinecite{parisi:96} by Parisi, Ricci-Tersenghi and 
Ruiz-Lorenzo, 
MPR Ref.~\onlinecite{marinari:98} by Marinari, Parisi, and Ruiz-Lorenzo, 
CEA Ref.~\onlinecite{campbell:00} by Campbell {\it et al}.,
and BY Ref.~\onlinecite{bhatt:88} by Bhatt and Young.
\label{crit_table_ISG}
}
\begin{tabular*}{\columnwidth}{@{\extracolsep{\fill}} c l c c c c }
\hline
\hline
$d$ & Reference & $T_{\rm c}$ & $\eta$ & $\nu$ & $z$ \\
\hline
$4$ & PRR       & $1.80(1)$   & $-0.35(5)$ & $1.0(1)$   &            \\
$4$ & CEA       & $1.78(1)$   & $-0.44(2)$ & $1.08(10)$ & $4.9(4)$   \\
\hline
$3$ & MPR       & $0.95(4)$   & $-0.36(6)$ & $2.00(15)$ &            \\
$3$ & CEA       & $0.92(2)$   & $-0.42(3)$ & $1.65(5)$  & $6.45(10)$ \\
\hline
$2$ & BY        & $0$         & $0$        & $3.63(10)$ &            \\
\hline
\hline
\end{tabular*}
\end{table}

The corrections to scaling in the GG follow just the same pattern as in
the bimodal ($\pm J$) ISGs. In the Gaussian ISG the corrections to scaling for
$\chi(L)$ at $T_{\rm c}$ appear very weak and so unmeasurable.  In dimension
four for both cases the correction to scaling for the susceptibility $\chi(L)$
is so weak as to be unobservable even down to $L=2$. In dimension three there
is a clear correction to scaling for $\chi(L)$ which can be fitted
satisfactorily to the form $\sim (1 + A L^{-\omega_{\rm eff}})$.
$\omega_{\rm eff}$ in the ISG case has been interpreted as the leading
irrelevant operator correction,\cite{mari:02} and this could also be the
case for the three-dimensional GG. However, it seems more likely that the
correction is dominated by a ``lattice artifact'' giving $\omega_{\rm eff}
= 2-\eta$, which has no relation to the leading irrelevant
operator.\cite{salas:98,salas:00} For both the three-dimensional GG and
the three-dimensional bimodal ISG, the value of the apparent correction to
scaling exponent is compatible with this interpretation, which would imply
that the leading irrelevant operator term has an exponent value $\omega >
2-\eta$.

Finally, we can give an overview of the critical exponents in the GG and
ISG families as a function of space dimension.  It should be noted that
there are some constraints that apply to both families: at the upper
critical dimension $d=6$, $\nu=1/2$, $\eta=0$, and $z=4$. If $T_{\rm c} = 0$,
$\nu=-1/\theta$, where $\theta$ is the stiffness exponent. Therefore, at
the lower critical dimension where the stiffness exponent $\theta=0$,
$\nu$ is infinite. For systems with a nondegenerate ground state (such as
the GG or the ISG with Gaussian interactions, but not the ISG with bimodal 
interactions), when $T=0$ there is no decay of the correlation function
$G(r)$ with $r$. This necessarily implies that when $T_{\rm c}=0$, $\eta(d)
\equiv 2-d$. [For systems with degenerate ground states $\eta(d) > 2 -
d$].

\begin{figure}
\centerline{\epsfxsize=\columnwidth \epsfbox{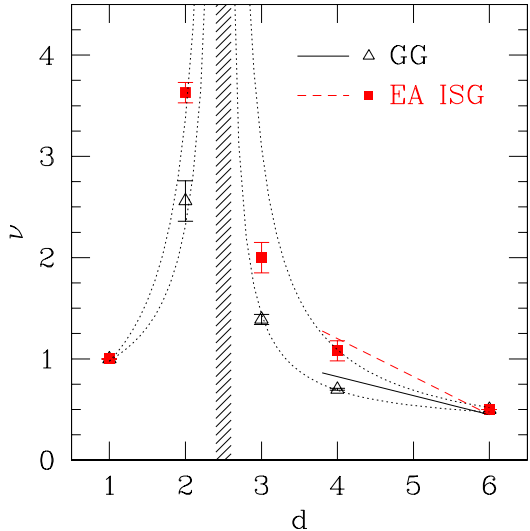}}
\vspace{-1.0cm}
\caption{(Color online)
Critical exponent $\nu$ as a function of space dimension for the ISG with
Gaussian bonds and the GG (data taken from Tables \ref{crit_table_ISG} and
\ref{crit_table}, respectively). The shaded region denotes the area where
one expects the lower critical dimension. The solid [dashed] line represents
the RG $\epsilon$-expansion estimate for the GG [ISG], as shown in the text,
and the dotted lines are guides to the eye. At the lower critical dimension 
one expects $\nu(d \rightarrow d_{\rm lcd}) \rightarrow \infty$. The data 
for the ISG and GG support this behavior.
}
\label{exp-nu_fig}
\end{figure}

\begin{figure}
\centerline{\epsfxsize=\columnwidth \epsfbox{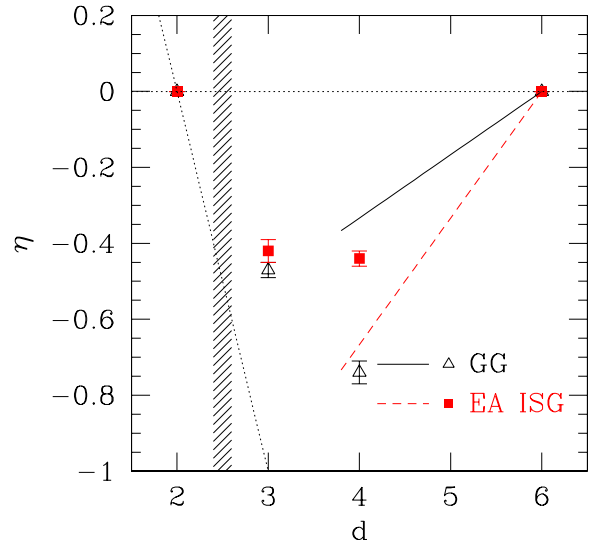}}
\vspace{-1.0cm}
\caption{(Color online)
Critical exponent $\eta$ as a function of space dimension. The
diagonal dotted line represents the physical limit line $\eta(d) = 2 - d$,
whereas the horizontal dotted line represents $\eta = 0$. The solid and 
dashed lines represent the $\epsilon$-expansion estimates for $\eta$ for 
the GG and ISG, respectively. For all $d < d_{\rm lcd} \approx 2.5$ $\eta(d)$ 
must join the limit line. The shaded region denotes the area where
one expects the lower critical dimension. 
}
\label{exp-eta_fig}
\end{figure}

The leading terms in the RG $\epsilon$-expansion for the GG are $\nu(d)=
1/2+ 5\epsilon/24$ and $\eta(d)=-\epsilon/6$. For the ISG $\nu(d)= 1/2+
5\epsilon/12$ and $\eta(d)=-\epsilon/3$.\cite{green:85,moore:90} It can
be seen in Figs.~\ref{exp-nu_fig} and \ref{exp-eta_fig}, respectively,
that in the GG as well as for the ISG the leading $\epsilon$-expansion
term gives the right sign and a qualitative indication of the strength of
the variations of the exponents just below the upper critical dimension.
The leading terms in the expansion for $\eta$ cannot, however, be used to
predict that the $\eta(d)$ values are much more negative in the GG than in
the ISG. The observed exponents deviate strongly from the curves
calculated to third order for the ISG where the second and third order
expansion terms are very large. This deviation is smaller for the GG. The
leading term in the $\epsilon$-expansion for the exponent $\nu$ works
better, with $\nu(d)$ being higher for the GG than for the ISG, in
agreement with the relative strengths of the leading terms (including
higher order terms would entirely destroy the agreement).  This is all in
striking contrast to the canonical ferromagnets (Ising, $XY$, or Heisenberg)
without disorder, where the third order $\epsilon$-expansion correction
gives excellent predictions for $\eta(d)$ an $\nu(d)$.

If we look at the overall behavior of the exponents over the whole range
of $d$, the shape of the $\nu(d)$ curves is fairly similar for the GG and
Gaussian ISG families, although the divergence at the lower critical
dimension $d_{\rm lcd} \approx 2.5$ is distinctly narrower in the GG case.  
The shapes of the $\eta(d)$ curves are on the other hand dramatically different.
The leading terms in the $\epsilon$-expansions agree poorly with the data
even for $d = 4$, with an inversion of the measured positions of $\eta(d)$ 
as compared with the expansion predictions.
For the Gaussian ISG family $\eta(d)$ decreases as the dimension drops from
the upper critical dimension to the lower critical dimension. For the GG
there is a deep minimum in $\eta(d)$ somewhere in the region of $d=4$
(data at $d=5$ would be needed to pin down the position of the minimum).
In spin glasses, simply going from Ising to vector spins changes this
exponent considerably, whereas in three-dimensional ferromagnets $\eta$ is
very small and practically independent of the type of spin symmetry.

\begin{figure}
\centerline{\epsfxsize=\columnwidth \epsfbox{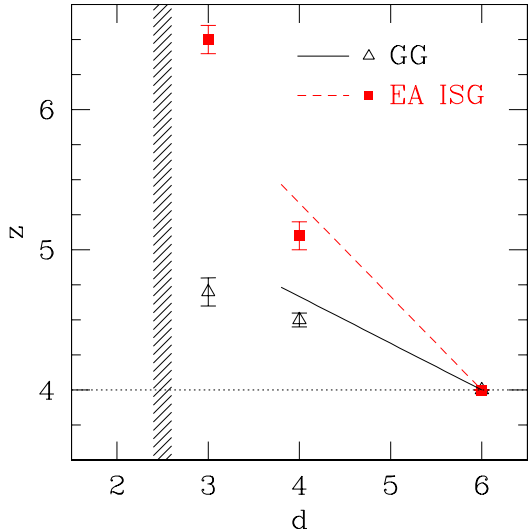}}
\vspace{-1.0cm}
\caption{(Color online)
Critical exponent $z$ as a function of space dimension. The horizontal
dotted line marks the mean-field asymptotic value $z = 4$, whereas the
shaded region denotes the area where one expects the lower critical
dimension. The solid (dashed) curve shows the asymptotic behavior as 
predicted from the van Hove approximation (Ref.~\onlinecite{zippelius:84}) 
for the GG (ISG). One expects that $z(d \rightarrow d_{\rm lcd}) \rightarrow \infty$. 
The ISG data show this behavior whereas the GG data are inconclusive.
}
\label{exp-z_fig}
\end{figure}

Finally, in Fig.~\ref{exp-z_fig} the critical dynamical exponent 
$z(d)$ can be compared near $d=6$ with the van Hove approximation
$z=2(2-\eta)$,\cite{zippelius:84} or for the ISG case with the first order
$\epsilon$-expansion $z=2(2-\eta)/(1+\epsilon/4)$.\cite{parisi:97} The
van Hove expression can provide a qualitative indication for $z(d)$ once
the numerical values of $\eta(d)$ are known, but if the directly measured
$\eta(d)$ values are used rather than the epsilon expansion estimates, 
this expression would predict higher $z$
values at each $d$ for the GG compared with the ISG values, which is not
what is observed. $z(d)$ must diverge at the lower critical dimension,
while curiously the GG $z(d)$ values only increase slightly between $d=4$
and $d=3$. For the Gaussian ISG, $z(d)$ shows an indication of divergence
as the space dimension drops.

\section{Concluding remarks}
\label{summary}

We have studied the critical parameters of the GG, in dimensions two, three and
four. The data confirm that $T_{\rm c}$ is zero in two dimensions.  It appears 
that the most reliable and accurate methods for estimating critical temperature
and exponent values are direct susceptibility measurements, effective
stiffness measurements (current measurements in the GG case), and a
combination of nonequilibrium and susceptibility measurements. In each
case corrections to scaling are either so small as to be negligible or can
be taken into account. Binder parameter measurements are inappropriate as
the Binder parameter curves do not intersect, at least for the system
sizes used, and correlation length measurements are not always reliable
because there can be strong deviations from finite-size scaling.

From these results a general strategy for obtaining (high-precision)
critical parameters numerically in any glassy system can be sketched out.
It is essential to be able to rely on high-quality equilibrium
susceptibility data over a wide range of sizes and to as large a size as
computational resources allow. Careful analysis of these data allowing for
the leading correction to scaling term should provide reliable estimates
of the ordering temperature and of the equilibrium critical exponents.  
Further measurements of the effective stiffness as a function of
temperature together with nonequilibrium behavior can then confirm the
value of the ordering temperature and give further information on other
exponents including the critical dynamical exponent $z$. Parameters
involving ratios (such as the Binder parameter or the correlation length)
seem frequently to be biased, at small and moderate sizes, by complicated
corrections to scaling and should be treated with caution.

If we compare with other systems having $d=6$ as upper critical dimension,
ISGs and percolation, the exponent $\nu(d)$ always evolves regularly,
diverging when the dimension reaches the lower critical dimension. The
exponent $\eta(d)$ on the other hand changes dramatically in form from one
family to the next. For the Gaussian ISG $\eta(d)$ grows smoothly more
negative as the dimension drops from the upper critical dimension to the
lower critical dimension. For the percolation systems $\eta(d)$ has a weak
minimum around $d=3$ before becoming positive at $d=2$.\cite{adler:90}
The present results show that for the GG there is a deep minimum in
$\eta(d)$ near $d=4$.  This is in stark contrast to the situation in the
canonical Ising, $XY$, or Heisenberg systems with no disorder where
$\eta(d)$ hardly changes at all when the degrees of freedom of the spin
are modified. In addition, for the canonical systems the RG
$\epsilon$-expansion to third order gives excellent predictions for the
exponents at dimensions well below the upper critical dimension, while for
the ISG and GG systems the $\epsilon$-expansion to the same
order\cite{green:85,moore:90} gives poor predictions for dimensions quite
close to the upper critical dimension $d=6$.

We have thus proposed a road plan for determining critical exponents
reliably and accurately from simulations; however, the results confirm once
again that the canonical RG theory lacks essential ingredients when
applied to glassy systems. The physical significance of the exponent
values obtained, however good they are, must remain obscure until the
appropriate theoretical approach going beyond the traditional RG is found
for interpreting them.

\begin{acknowledgments}

We would like to thank J.~Salas for helpful correspondence. C.~De
Dominicis and A.~J.~Bray kindly supplied us with the leading term for the
exponent $\omega$ in the $\epsilon$-expansion for the ISG and the GG
cases, respectively. The simulations were performed on the Asgard cluster
at ETH Z\"urich.

\end{acknowledgments}

\bibliography{refs,comments}

\begin{thebibliography}{57}
\bibitem[{cur()}]{curve}
\bibinfo{note}{{This effect is weak and can only be seen in a full-size plot.}}
\bibitem[{exp()}]{exponents}
\bibinfo{note}{{Note that for the upper critical dimension, $d = 6$, we expect
  mean-field exponents, i.e., $\nu = 1/2$, $\eta = 0$, and $z = 4$.}}
\end{thebibliography}

\end{document}